\documentclass[twocolumn]{aastex631}

\usepackage{natbib}
\usepackage{epsfig}
\usepackage{graphicx}
\usepackage{subfigure}
\usepackage{float}
\usepackage{amsmath}
\usepackage{color}
\usepackage{amssymb}
\usepackage{apjfonts}

\received{September 6, 2024}

\shortauthors{Liu \& Melia}

\graphicspath{{./}{figures/}}

\usepackage{amsmath}
\begin{document}

\title{A Truncated Primordial Power Spectrum and \\ its Impact on CMB Polarization}


\correspondingauthor{Fulvio Melia}
\email{fmelia@email.arizona.edu}

\author{Jingwei Liu}
\affiliation{Department of Physics, The University of Arizona, AZ 85721, USA}

\author{Fulvio Melia \thanks{John Woodruff Simpson Fellow.}}
\affiliation{Department of Physics, The Applied Math Program and Department of Astronomy,
The University of Arizona, AZ 85721, USA}

\begin{abstract}
We investigate the impact of a hypothesized delayed initiation of inflation, 
characterized by a cutoff $k_{\rm min}$ to the primordial power spectrum in the 
cosmic microwave background (CMB). This cutoff affects both the scalar and tensor 
spectra, which therefore impacts several measurements of the temperature and 
polarization distributions. We calculate the angular power spectrum and correlation 
function with and without $k_{\rm min}$ in the context of {\it Planck}-$\Lambda$CDM, 
and demonstrate that a non-zero $k_{\rm min}$ significantly improves the alignment 
between theory and the observations, including the temperature, $E$-mode polarization, 
$TE$ cross-correlation, $Q+U$ polarization and $Q-U$ polarization. It creates an
observable signature in both the angular power spectrum and correlation function 
for all cases. We thus also explore the $B$-mode polarization, for which current
data are not yet precise enough to determine $k_{\rm min}$, but whose impact should 
be detectable with high-precision measurements using future missions, such as LiteBIRD, 
if the tensor-to-scalar ratio, $r$, is not much smaller than its current upper limit. 
We find that the introduction of $k_{\rm min}$ not only addresses large-angle anomalies 
in the CMB but also provides a more consistent framework for understanding the early 
Universe's inflationary phase. These findings highlight the importance of future 
high-precision CMB observations in validating the existence and implications of 
$k_{\text{min}}$.
\end{abstract}


\section{Introduction}\label{intro}
Inflation was introduced over four decades ago \citep{Starobinskii:1979,Kazanas:1980,Guth:1981,Linde:1982}
to solve several major problems with standard cosmology, including an inexplicably uniform 
temperature in the cosmic microwave background (CMB) across causally disconnected regions of the sky, 
flatness \citep{Melia:2022a} and the absence of magnetic monopoles that should have been produced in 
large quantities in the context of grand unified theories (GUTs) \citep{Melia:2023f}. We still do not 
know the underlying physics of inflation, however, and very little information has been extracted
thus far from the observations concerning its initiation and a possible pre-inflationary epoch.

More recently, largely due to the success of the {\it Planck} satellite \citep{PlanckVI:2020}, 
unprecedented progress has been made with observations of the CMB but, contrary to expectations, 
inflationary theory has not been clearly confirmed. Instead, the traditional cosmological model 
is facing an increasing number of challenges. For example, while inflation predicts a nearly 
scale-invariant spectrum of primordial perturbations \citep{Mukhanov:1992}, certain anomalies 
in the CMB---such as a lack of large-angle correlations \citep{Copi:2010}, a hemispherical 
asymmetry \citep{Eriksen:2004} and a `cold' spot \citep{Cruz:2005}---are not fully explained by 
standard inflationary models.  Moreover, the theory itself was not 
established with the full compliance of General Relativity (GR)
and Quantum Field Theory (QFT), so its internal structure is fragile in many ways. As a result, an 
increasing number of challenges to the core theory have been raised over the past few decades. Most 
prominently, questions have been asked concerning its initial conditions 
\citep{Ijjas:2013,Ijjas:2014}, its informal (and possibly inconsistent) handling of quantum 
fluctuations in the inflaton field \citep{Martin:2001} and its evident violation of the strong 
energy condition in GR \citep{Melia:2023e}.

The most serious concern regarding inflation is its lack of predictive power. Many efforts have 
been made over the years to mitigate at least some of the observational anomalies by modifying 
the basic model. These efforts have led to an accommodation with the data, but at the expense of
increasing the model's flexibility, which also makes it difficult to falsify, since the model
can easily be adjusted to fit essentially `anything.'

Over the coming years, one of the most important observational developments related to inflation
will be an attempt to measure B-mode polarization in the CMB fluctuations, believed to represent 
a tell-tale signature of tensor perturbations in the inflaton field \citep{Mukhanov:1992}. 
While other mechanisms may also create such quantum fluctuations, it is understood that an 
absence of the signature they would leave behind in the CMB temperature profile would argue 
strongly against inflation having actually happened. Our focus in this paper will thus be an 
exploration of how $k_{\rm min}$ could impact the CMB polarization signal, the $E$-mode part 
of which is already measured by {\it Planck} \citep{PlanckVI:2020} and the $B$-mode part to 
be measured, e.g., by LiteBIRD \citep{LiteBIRD:2022}.

We take as our starting point the anomalous lack of large-angle correlations mentioned earlier,
which appears to be due to a wavenumber cutoff, $k_{\rm min}$, in the primordial power spectrum 
\citep{MeliaLopez:2018,Melia:2021b,Sanchis-Lozano:2022}. In so doing, however, we shall attempt
to adhere as much as possible to the conventional picture of inflation, incorporated into the
{\it Planck}-$\Lambda$CDM (standard) model. As we shall explain below, the introduction of
$k_{\rm min}$ is strongly motivated by the physics of the early Universe 
\citep{LiuMelia:2020,Liu:2024a}. And such a cutoff is actually required by the initiation 
of a slow-roll inflationary phase.

We begin our discussion by briefly summarizing the basic picture of how the temperature 
and polarization fluctuations in the CMB are defined and how they are thought to arise
from the primordial quantum fluctuations. We shall focus on two primary objectives. First, 
we shall discuss why polarization modes are important, and how $B$ and $E$ modes are defined, 
distinguished and used. Second, we shall elaborate upon the notion that $k_{\rm min}$ 
corresponds to the initiation of a slow-roll inflationary expansion and how it impacts 
our understanding of how quantum fluctuations are generated and grown. 

After this brief background, we shall explore the differences between models with and 
without $k_{\rm min}$, notably in their predicted angular correlation function and the 
corresponding angular power spectrum. We shall quantify the impact of $k_{\rm min}$ on
these observables, and compare the results to the existing measurements from {\it Planck}
\citep{PlanckVI:2020} and examine the likelihood of them being measured by upcoming
missions, such as LiteBIRD \citep{LiteBIRD:2022}. 

\section{Calculation of the angular power spectrum and angular correlation function}\label{angpower}
\subsection{Brief Theoretical Background}\label{theo}
Like any radiation field, the CMB can be characterized by the intensity tensor 
$I_{ij} $, from which we obtain the four well-known Stokes parameters:
\begin{equation}
\begin{aligned}
T &= \frac{I_{11} + I_{22}}{4}\,,\\
Q &= \frac{I_{11} - I_{22}}{4}\,,\\
U &= \frac{I_{12} + I_{21}}{4}\,,\\
V &= \frac{I_{12} - I_{21}}{4i}\,.
\end{aligned}
\label{stokes_parameters}
\end{equation}
For CMB studies, only the first three of these quantities are important because Thomson 
scattering does not generate the $V$ component of polarization \cite{Zaldarriaga:1997}.

We can directly use $T$, $Q$, and $U$ to analyze the CMB, especially if we only focus on 
a small fraction of the sky, which can be approximated as a (local) plane. If the analysis 
involves large separations on the sky, however, $Q$ and $U$ are not appropriate because their
evaluation relies on choosing a specific direction as the reference. But such a direction
is not available on a sphere, which one would need to use for a large fraction of the sky. 
Therefore, we need to introduce the so-called $B$ and $E$ modes, which are rotationally 
independent, for the analysis of the CMB polarization \citep{Zaldarriaga:1997,Yoho:2014}. 
To see how these two sets of parameters are related, let us first examine the properties 
of $Q$ and $U$. 

By rotating the sky by an angle $\phi$, one gets \citep{Kosowsky:1996}
\begin{equation}
\begin{aligned}
Q' &= Q\cos(2\phi) + U\sin(2\phi)\,,\\
U' &= -Q\sin(2\phi) + U\cos(2\phi)\,,
\end{aligned}
\label{QU_rotate}
\end{equation}
which may be written more compactly as 
\begin{equation}
\begin{aligned}
(Q \pm i U)' = e^{\mp 2i \phi}(Q \pm i U)\,.
\end{aligned}
\label{Q+-iU}
\end{equation}
It is therefore common to construct a pair of spin-weighted functions, $Q+iU $ and 
$Q-iU $, with a spin weight of 2. Notice that $T $ is already rotation invariant, 
meaning it is a spin-weighted function with a spin weight of 0. Therefore, one may
expand $T $, $Q $, and $U $ in terms of spin-weighted spherical harmonics 
\citep{Zaldarriaga:1997}:
\begin{equation}
\begin{aligned}
    T &= \sum_{\ell,m} a_{T,\ell m} Y_{\ell m}\,, \\
    Q+iU &= \sum_{\ell,m} a_{2,\ell m} {}_{2}Y_{\ell m}\,, \\
    Q-iU &= \sum_{\ell,m} a_{-2,\ell m} {}_{-2}Y_{\ell m}\,.
\end{aligned}
\label{SphericalExpandedTQU}
\end{equation}

The benefit of doing this is that, for spin-weighted functions, we can apply spin raising 
and lowering operators to transform them into spin-0 functions, which are then 
rotation-invariant. This step removes any ambiguity that arises from the definition
of a reference direction in the sky when calculating the $Q$ and $U$ modes. The operation 
proceeds as follows \citep{Zaldarriaga:1997}: 
\begin{equation}
\begin{aligned}
\bar{\eth}^2 (Q+iU) &= \sum_{\ell,m} \left( \frac{(\ell+2)!}{(\ell-2)!} 
	\right)^{1/2} a_{2,\ell m} Y_{\ell m}\,, \\
\eth^2 (Q-iU) &= \sum_{\ell,m} \left( \frac{(\ell+2)!}{(\ell-2)!} 
	\right)^{1/2} a_{-2,\ell m} Y_{\ell m}\,.
\end{aligned}
\label{Expanded_Q+-iU}
\end{equation}
The better known $B$ and $E$ modes are then defined as
\begin{equation}
\begin{aligned}
a_{E,\ell m} &= -\frac{a_{2,\ell m} + a_{-2,\ell m}}{2}\,, \\
a_{B,\ell m} &= i \frac{a_{2,\ell m} - a_{-2,\ell m}}{2}\,, \\
E &= \sum_{\ell m} a_{E,\ell m} Y_{\ell m}\,, \\
B &= \sum_{\ell m} a_{B,\ell m} Y_{\ell m}\,.
\end{aligned}
\label{Defof_E_B}
\end{equation}

To explain how the perturbations in the CMB were generated, we adopt the synchronous 
gauge, in which a metric perturbation is written as \citep{Ma:1995}
\begin{equation}
ds^2 = a^2(\tau)\left\{-d\tau^2 + (\delta_{ij} + h_{ij})dx^i dx^j\right\}\;.
\label{DefofSynch}
\end{equation}
And in this framework, we introduce two fields in $k$-space, $h(k, \tau)$ and $\eta(k, \tau)$,
such that
\begin{equation}
\begin{aligned}
h_{ij}(x, \tau) &= \int d^3k \, e^{ikx} (\hat{k}_i \hat{k}_j h(k, \tau) \\
&+ \left(\hat{k}_i \hat{k}_j - \frac{1}{3}\delta_{ij} ) 6\eta(k, \tau)\right)\;.
\end{aligned}
\label{h_i_j}
\end{equation}

The scalar mode perturbations in the CMB may then be evolved according to the
Boltzmann equations in the synchronous gauge as follows \citep{Zaldarriaga:1997,Kosowsky:1996}:
\begin{equation}
\begin{aligned}
\dot{\Delta}^{(S)}_T + ik\mu \Delta^{(S)}_T &= -\frac{1}{6} \dot{h} - \frac{1}{6} (\dot{h} + 
6\dot{\eta}) P_2(\mu) + \dot{\kappa} \\
&- \Delta^{(S)}_T + \Delta^{(S)}_{T0} + i\mu v_b + \frac{1}{2} P_2(\mu) \Pi \,, \\
\dot{\Delta}^{(S)}_P + ik\mu \Delta^{(S)}_P &= \dot{\kappa} - \Delta^{(S)}_P + \frac{1}{2} [1 - P_2(\mu)] \Pi \,,
\end{aligned}
\label{Boltz_Scalar}
\end{equation}
\begin{equation}
\Pi = \Delta^{(S)}_{T2} + \Delta^{(S)}_{P2} + \Delta^{(S)}_{P0} \,.
\label{Def_Pi}
\end{equation}
In these expressions, $\Delta^{(S)}_T$ represents the temperature anisotropy, and a specific 
direction is chosen for the definition of polarization. For a given Fourier mode, the coordinate 
system is selected such that the wave vector $k$ is parallel to the $\hat{z}$-direction. Here, 
$\mu$ is defined as $\hat{n} \cdot \hat{k}$, where $\hat{n}$ denotes the direction of the photon. 
The $Q$-mode of the polarization in this coordinate system is denoted as $\Delta^{(S)}_P$, and 
there is no $U$-mode for such a coordinate.

Additional parameters include: $v_b$, which stands for the bulk baryon velocity, and 
$\dot{\kappa}$, which represents the differential optical depth and can be calculated 
from $\dot{\kappa} = a(\tau) n_e x_e \sigma_T$, where $a(\tau)$ is the scale factor, $n_e$ is the electron 
density, $x_e$ is the ionization fraction, and $\sigma_T$ is the Thomson cross-section.

The definition of $\Pi$ requires further explanation. It represents the combination of 
multiple moments of $\Delta^{(S)}_P$ and $\Delta^{(S)}_T$, as shown in Equation~(\ref{Def_Pi}). 
The multiple moments are defined in terms of the Legendre expansion of $\Delta^{(S)}_P$ and 
$\Delta^{(S)}_T$, based on the general expression
\begin{equation}
\Delta = \sum_{\ell}(2\ell+1)(-i)^\ell\Delta_\ell(k)P_\ell(\mu)\;.
\label{Def_MultipleMoments}
\end{equation}
To obtain the perturbations we see today, one integrates the Boltzmann equations from the 
last scattering surface to today along the line of sight, yielding the results 
\citep{Seljak:1996,Zaldarriaga:1997}
\begin{equation}
\begin{aligned}
\Delta^{(S)}_T (\tau_0, k, \mu) &= \int_0^{\tau_0} d\tau \, e^{ix\mu} S^{(S)}_T (k, \tau) \,, \\
\Delta^{(S)}_P (\tau_0, k, \mu) &= \frac{3}{4} (1 - \mu^2) \int_0^{\tau_0} d\tau \, 
	e^{ix\mu} g(\tau) \Pi(k, \tau) \,, \\
S^{(S)}_T (k, \tau) &= g \left[ \Delta_{T,0} + 2\dot{\alpha} + \frac{\dot{v_b}}{k} + 
	\frac{\Pi}{4} + \frac{3\ddot{\Pi}}{4k^2} \right] \\
&+ e^{-\kappa} (\dot{\eta} + \ddot{\alpha}) + \dot{g} \left[ \alpha + \frac{v_b}{k} + 
	\frac{3\dot{\Pi}}{4k^2} \right] \\
&+ \frac{3\ddot{g}\Pi}{4k^2}\,. \\
\end{aligned}
\label{Int_Boltz_Scalar}
\end{equation}

In these expressions, we have introduced several new variables, including $x\equiv k(\tau_0 - \tau)$, 
where $\tau_0$ is defined today, while $\tau$ is at the last scattering surface. Additionally, 
$\alpha \equiv (\dot{h} + 6\dot{\eta})/{2k^2}$, and $g(\tau) \equiv \dot{\kappa} \exp(-\kappa)$.

The $Q$-mode of polarization may be converted into the $E$-mode and $B$-mode using the spin 
raising and lowering operators introduced earlier. For scalar modes, the $B$-mode vanishes 
automatically, leaving only the $E$-mode. Then, Equation~(\ref{Int_Boltz_Scalar}) becomes
\begin{equation}
\begin{aligned}
\Delta^{(S)}_{T\ell}(k) &= \int_{0}^{\tau_0} d\tau S^{(S)}_T(k, \tau) j_\ell(x)\,, \\
\Delta^{(S)}_{E\ell}(k) &= \sqrt{\frac{(l+2)!}{(l-2)!}} \int_{0}^{\tau_0} 
	d\tau S^{(S)}_E(k, \tau) j_\ell(x)\,,
\end{aligned}
\label{Int_Boltz_Scalar_TE}
\end{equation}
in which 
\begin{equation}
 S^{(S)}_E(k, \tau) = \frac{3g(\tau)\Pi(\tau, k)}{4x^2}
\label{Source_Scalar_E}
\end{equation}
is the source term of the $E$-mode.

The power spectra for T and E and their cross-correlation are simply given by
\begin{equation}
\begin{aligned}
C^{(S)}_{T,E \ell} &= (4\pi)^2 \int k^2 dk \, P_\phi(k) \left[ \Delta^{(S)}_{T,E \ell}(k) \right]^2 \,,\\
C^{(S)}_{C \ell} &= (4\pi)^2 \int k^2 dk \, P_\phi(k) \Delta^{(S)}_{T \ell}(k) \Delta^{(S)}_{E \ell}(k)\,,
\end{aligned}
\label{Powerspectra_Scalar}
\end{equation}
in which $P_\phi(k)$ stands for the primordial power spectrum of the scalar mode.
We thus see that the angular power spectrum generated by the scalar modes can be 
calculated by solving the set of differential Equations~(\ref{Def_Pi}), (\ref{Int_Boltz_Scalar}), 
(\ref{Int_Boltz_Scalar_TE}) and (\ref{Source_Scalar_E}).

Analogously to the scalar modes, we can derive a set of differential equations for the tensor
modes based on the Boltzmann equations. These are 
\begin{equation}
\begin{aligned}
\widetilde{\Delta}^{(T)}_T (\tau, \mu, k)
&=\int_{0}^{\tau} d\tau e^{ix\mu} S^{(T)}_T (k, \tau) \,,\\
\widetilde{\Delta}^{(T)}_P (\tau, \mu, k)
&=\int_{0}^{\tau} d\tau e^{ix\mu} S^{(T)}_P (k, \tau)\,,
\end{aligned}
\label{Int_Boltz_tensor}
\end{equation}
\begin{equation}
\begin{aligned}
S^{(T)}_T (k, \tau) &= -\dot{h} e^{-\kappa} + g \Psi \,, \\
S^{(T)}_P (k, \tau) &= -g \Psi\,,
\end{aligned}
\label{Source_Tensor_TP}
\end{equation}
and
\begin{equation}
\begin{aligned}
\Psi &\equiv ( \frac{1}{10} \widetilde{\Delta}^{(T)}_{T0} + \frac{1}{7} 
	\widetilde{\Delta}^{(T)}_{T2} + \frac{3}{70} \widetilde{\Delta}^{(T)}_{T4} \\
&-\frac{3}{5} \widetilde{\Delta}^{(T)}_{P0} + \frac{6}{7} 
	\widetilde{\Delta}^{(T)}_{P2} - \frac{3}{70} \widetilde{\Delta}^{(T)}_{P4} )\,.
\end{aligned}
\label{Def_Psi}
\end{equation}

Combining these equations, we may solve for $\Psi$ at any given $\tau$. In addition,
\begin{equation}
\begin{aligned}
\Delta^{(T)}_{Tl} &= \sqrt{\frac{(l+2)!}{(l-2)!}} \int_0^{\tau_0} d\tau \, 
	S^{(T)}_T(k,\tau) \frac{j_l(x)}{x^2} \,,\\
\Delta^{(T)}_{E,Bl} &= \int_0^{\tau_0} d\tau \, S^{(T)}_{E,B}(k,\tau) j_l(x)\,,
\end{aligned}
\label{Int_Boltz_Tensor_TEB}
\end{equation}
and
\begin{equation}
\begin{aligned}
S^{(T)}_E(k,\tau) &= g \left( \Psi - \frac{\ddot{\Psi}}{k^2} + \frac{2 \Psi}{x^2} - 
	\frac{\dot{\Psi}}{kx} \right) \\
&- \dot{g} \left( \frac{2 \dot{\Psi}}{k^2} + \frac{4 \Psi}{kx} \right) - 
	\frac{2 \ddot{g} \Psi}{k^2} \,,\\
S^{(T)}_B(k,\tau) &= g \left( \frac{4 \Psi}{x} + \frac{2 \dot{\Psi}}{k} \right) + 
	\dot{g} \left( \frac{2 \dot{\Psi}}{k} \right)\,.
\end{aligned}
\label{Source_Tensor_BE}
\end{equation}

The angular power spectra are then found from the expressions \citep{Zaldarriaga:1997} 
\begin{equation}
\begin{aligned}
C^{(T)}_{Xl} &= (4\pi)^2 \int k^2 dk \, P_h(k) \left[ \Delta^{(T)}_{Xl}(k) \right]^2 \,,\\
C^{(T)}_{Cl} &= (4\pi)^2 \int k^2 dk \, P_h(k) \Delta^{(T)}_{Tl}(k) \Delta^{(T)}_{El}(k)\,,
\end{aligned}
\label{PowerSpectra_Tensor}
\end{equation}
in which $X$ stands for $T$, $E$ or $B$. Thus we can solve for $\Psi$ from 
Equations~(\ref{Int_Boltz_tensor})--(\ref{Def_Psi}), and then use 
Equations~(\ref{Int_Boltz_Tensor_TEB}) and (\ref{Source_Tensor_BE}) to find the 
angular power spectrum.

Now let us briefly review how the observable fluctuations were generated from the 
nearly scale-free primordial spectrum, which itself was created at horizon crossing. 
The core concept is that, during inflation, the Universe expanded at close to
an exponential rate, while the Hubble radius 
\begin{equation}
R_H = \frac{c}{H}
\label{R_H}
\end{equation}
remained approximately constant. The quantum fluctuation wavelengths, however,
continued to grow at a rate proportional to the expansion factor, $a(t)$: 
\begin{equation}
\lambda(k) = a \frac{2\pi}{k}\;.
\label{Lambda_k}
\end{equation}

Thus, modes that were initially smaller than $R_H$ grew larger than the Hubble radius 
and left the horizon. The smaller $k$ is (i.e., the larger the comoving wavelength), 
the earlier the mode crossed, satisfying the condition \citep{Liddle:1994}
\begin{equation}
a H=c k\label{across} \,.
\end{equation}

With this conventional picture, it is easy to see
why a hard cutoff $k_{\rm min}$ in the primordial power spectrum should correspond 
to the initiation of inflation. If inflation started at a specific time, there must 
be a largest mode whose wavelength matches the Hubble radius at that instant. 
Therefore, $k_{\rm min}$ serves as a strong constraint one may use to determine 
the time at which inflation started \citep{LiuMelia:2020,Liu:2024a}.

Of course, the initial motivation for introducing $k_{\rm min}$ was more empirical,
essentially to mitigate the well-known large-angle anomalies in the CMB, confirmed 
by multiple observational missions, such as the Cosmic Background Explorer 
(COBE) \citep{Hinshaw:1996}, the Wilkinson Microwave Anisotropy Probe (WMAP) 
\citep{Bennett:2003} and the {\it Planck} mission \citep{Planck:2014,PlanckVI:2020}. 
The most significant finding of these is that the data exhibit a lack 
of any correlation in the temperature fluctuations on angular scales 
larger than about 60 degrees. A second anomaly is associated with the fact that
the measured power in the temperature angular power spectrum is significantly
lower than expected in the standard model at low $\ell$s ($\ell \leq 4$). Several
attempts have been made to mitigate this particular problem by modifying the 
primordial power spectrum or introducing some kind of pre-inflationary phase 
\citep{Contaldi:2003,Hazra:2014,Schwarz:2016}. 

The first application of a cutoff $k_{\text{min}}$ was to address the temperature 
angular correlation function \citep{MeliaLopez:2018}. This work showed that a zero 
$k_{\rm min}$ model is ruled out at a confidence level of over $8\sigma$. It also 
showed that the angular correlation function calculated with non-zero values of 
$k_{\rm min}$ fits the data much better than the basic standard model without
the cutoff---at all angles, not just above 60 degrees. The optimized value of 
$k_{\rm min}$ found from the analysis of $C_{TT}(\theta)$, which we shall be 
using for the rest of this paper, is 
\begin{equation}
k_{\rm min} = (3.14 \pm 0.36) \times 10^{-4} \quad \text{Mpc}^{-1}\;.
\label{kmin}
\end{equation}

To emphasize the significance of such a robust result, it is worth reminding
ourselves of the procedure followed to achieve the $8\sigma$ confidence level. 
The data points in the observed angular correlation function $C(\theta)$ are highly 
correlated, which must thus be considered in the statistical analysis. The optimized 
value of $k_{\rm min}$ is determined by analyzing the distribution of mock CMB 
catalogs, which utilize the standard cosmological model with the angular correlation 
function $C_{TT}(\theta)$. The uncertainty in this quantity is obtained from the 
r.m.s. value that encompasses $68\%$ of the possible outcomes. This 
error margin is larger than what one would derive from a simple $\chi^2$ fitting, 
which does not account for the correlations. Therefore, the $8\sigma$ estimate 
represents the most conservative limit one may adopt in the independent analysis 
of the angular correlation function.

Subsequent to this work, a second study with the same basic concept---that the 
large angle anomaly is due to a minimum $k_{\rm min}$---was carried out, though
this time focusing on the angular power spectrum \citep{Melia:2021b}. This analysis
also showed a great improvement in fitting the data at $\ell \leq 5$, with a cut-off 
$k_{\rm min} = (2.04^{+1.4}_{-0.79}) \times 10^{-4} \; {\rm Mpc}^{-1}$, optimized 
using solely the angular power spectrum rather than the angular correlation function. 
These two values of $k_{\rm min}$ are clearly fully consistent with each other
within $1\sigma$.

We need to emphasize that, though both papers \citep{MeliaLopez:2018,Melia:2021b} 
focused on large angle anomalies, 
the problems they addressed were two separate features of the CMB. The first paper 
focused on the temperature angular correlation function ,while the second addressed 
the low power seen at low $\ell$'s in the angular power spectrum. Also, it must be
recognized that, though $k_{\rm min} = 0$ is ruled out at a very high level of
confidence, it does not mean that the standard cosmological model itself is ruled 
out. It merely suggests that, to fit the data as optimally as possible, a non-zero 
$k_{\rm min}$ must be included. One of the more important consequences of this
empirical finding is that the existence of a non-zero $k_{\rm min}$ points to
a delayed initiation of slow-roll inflation \citep{LiuMelia:2020,Liu:2024a}.

To make this linkage between $k_{\rm min}$ and the initiation of inflation
more quantitative, we shall use one of the best known slow-roll models, 
based on a Higgs-like potential \citep{Bezrukov:2008,Barvinsky:2008,Lee:2018,Liu:2024a}, 
\begin{equation}
V=V_0\left[1-\left(\frac{\phi}{\mu}\right)^2\right]^2\,,\label{eq:Higgslike}
\end{equation}
from which we can derive the slow-roll parameters \citep{Kolb:1994},
\begin{equation}
\epsilon_V\equiv \frac{M_{\rm Pl}^2}  {2}\left(\frac{V_{,\phi}}  {V}\right)^2
=\frac{8M_{\rm Pl}^2\phi^2}{\mu^4[1-({\phi / \mu})^2]^2}\,,\label{eq:HiggsEpsilon}
\end{equation}
and
\begin{equation}
\eta_V\equiv M_{\rm Pl}^2\frac{V_{,\phi\phi}}{V}
=-\frac{4M_{\rm Pl}^2(\mu^2-3\phi^2)}{(\mu^2-\phi^2)^2}\,,\label{eq:HiggsEta}
\end{equation}
where $V_{,\phi}$ denotes the first derivative of $V$ with respect to $\phi$, and 
$V_{,\phi\phi}$ is its second derivative. In these expressions, $M_{\rm Pl}$ is 
the reduced Planck mass, defined as $M_{\rm Pl} = m_{\rm Pl}/\sqrt{8\pi}$.
Under the slow-roll approximation, we obtain:
\begin{equation}
n_{\rm s}-1 = 2\eta_V-6\epsilon_V\,,
\end{equation}
and
\begin{equation}
r = 16\epsilon_V\,.
\end{equation}
Adopting the {\it Planck} measurement, $n_{\rm s}=0.966$, and the upper limit of the 
tensor-to-scalar ratio, $r=0.036$ \citep{Ade:2021}, we find that $\mu \approx 18.7\;
M_{\rm Pl}$ and $\phi_{0.002} \approx 5.4\;M_{\rm Pl}$ (the value of $\phi$ at 
$k=0.002$ Mpc$^{-1}$). The fact that $\left({\phi_{0.002}}/{\mu}\right)^2\ll1$ 
confirms our inference that $V(\phi)$ would have been very nearly constant at the early 
stage of inflation. As long as $V(\phi)$ dominated the energy density of the Universe 
at that point, we conclude that $H$ must also have been approximately constant during 
the early stages of inflation. With slow-roll \citep{Linde:1983},
\begin{equation}
\Delta_{\rm s}^2 = \frac{H^2}{8 \pi^2 M_{\rm Pl}^2 \epsilon}=\frac{2H^2}{\pi^2 
	M_{\rm Pl}^2 r}\;.\label{eq:Deltak2}
\end{equation}
But from the definition of the amplitude, $A_{\rm s}$, of the primordial power spectrum, 
we also have
\begin{equation}
\Delta_{\rm s}^2 = A_{\rm s}\left(\frac{k}{k_*}\right)^{n_{\rm s}-1}\;,\label{eq:As}
\end{equation}
where $k_*$ is a pivot scale, usually taken to be $0.05$ Mpc$^{-1}$. Thus, inserting 
the {\it Planck} measured value of the amplitude, $A_{\rm s} = 2.1 \times 10^{-9}$, 
we conclude that $H_{\rm init} \approx H_{0.002}=3.1 \times 10^{37}$ s$^{-1}$. And, 
combining this with Equation~(\ref{across}) completely determines the initiation of 
inflation. From the procedure we just demonstrated, it is clear that the energy 
density and time at the beginning of slow-roll inflation are model-dependent. 
Nevertheless, it is clear how $k_{\rm min}$ corresponds to the initiation time
and how it plays a crucial role in constraining the cosmological model 
\citep{LiuMelia:2020,Liu:2024a}.

With this conclusion, one therefore needs to modify the expressions for 
the angular power spectrum and polarization of the CMB temperature as follows: 
\begin{equation}
\begin{aligned}
C^{(S)}_{T,E\ell} &= 4 \int_{k_{\text{min}}}^{\infty} dk \, k^2 P_\phi(k) 
	\left[ \Delta^{(S)}_{T,E\ell}(k) \right]^2 \,,\\
C^{(T)}_{T,E,B\ell} &= 4 \int_{k_{\text{min}}}^{\infty} dk \, k^2 P_h(k) 
	\left[ \Delta^{(T)}_{T,E,B\ell}(k) \right]^2 \,.
\end{aligned}
\label{AngulrPowerSpectra_withkmin}
\end{equation}

Clearly, $k_{\rm min}$ affects all the $C_\ell$'s at all $\ell$'s, and therefore 
affects $C(\theta)$ at all angles, giving us the opportunity of confirming the existence 
of $k_{\rm min}$ using the polarization data as well as the temperature angular power 
spectrum and angular correlation function. It is worth mentioning that the impact of 
$k_{\rm min}$ on the angular power spectrum is different than that on the angular 
correlation function. This happens because $k_{\rm min}$ is  a large-scale feature, 
so it mainly affects the large-scale end of the power spectrum. Its impact is felt 
mainly on the low $\ell$'s, as we shall show in the next section. In other words,
it suppresses the $C_{\ell}$ values at the low $\ell$'s, especially at $\ell \leq 4$. 
When it comes to the angular correlation function, however, every $C_{\ell}$ 
contributes to $C(\theta)$ at all angles, so $k_{\rm min}$'s impact is seen at
all $\ell$'s. 

The evidence for the existence of $k_{\rm min}$ provided by the temperature fluctuations 
in the CMB is already quite compelling \citep{MeliaLopez:2018,Melia:2021b,Liu:2024PLB}, 
but there are multiple reasons for delving deeper
into this problem by also analyzing its impact on the polarization of the CMB. 

First, 
there already exist some high-quality polarization data, thanks to the success of 
the {\it Planck} mission. Analyzing $k_{\rm min}$'s impact on these observations
can place our analysis on firmer ground.

Second, almost all of the existing and forthcoming CMB observational campaigns are 
aiming to measure the CMB's polarization more precisely. If the existing or forthcoming 
polarization data are precise enough to tell the difference between models with and without 
$k_{\rm min}$, they would provide an indispensable window into the initiation time
of inflation. Moreover, there is still some doubt that the large-angle anomalies represent
real physics, that they are somehow a statistical `fluke', but a strong confirmation of a 
non-zero $k_{\rm min}$ would completely rule out that possibility. 

Third and most importantly, if a non-zero $k_{\rm min}$ is indeed due to a delay of the 
initiation of inflation, then it should not only affect the scalar primordial power spectrum 
but the tensor power spectrum as well, whose analysis requires the polarization data, 
especially the $B$-mode polarization, which can only be generated by tensor modes (or 
more weakly through lensing effects). Thus, the observation of $B$-mode polarization
is crucial for confirming the existence of $k_{\rm min}$ and improving our understanding 
of the early Universe. 

The above discussion has summarized the theoretical motivation and procedure for
calculating the angular power spectrum and angular correlation function of both
the temperature fluctuations and the polarization of the CMB within the context
of standard cosmology. For the remainder of this paper, most of the calculations are
carried out using the CAMB code (version 1.4.0) \citep{Lewis:2000} with the most recent 
{\it Planck} results \citep{PlanckVI:2020} as its default parameters. To incorporate 
the impact of $k_{\rm min}$, we do not use CAMB's default primordial power spectrum. 
Instead, the modified primordial power spectrum used for the scalar modes is 
\begin{equation}
P^{(S)}(k) = 
\begin{cases} 
A_{\rm s} \cdot 
\left(\frac{k}{k_{\rm pivot}}\right)^{n_{\rm s} - 1} & \text{for } k \geq k_{\rm min}\,.
\end{cases}
\label{Primordial_Scalar}
\end{equation}
For the tensor modes, we use 
\begin{equation}
P^{(T)}(k) = 
\begin{cases} 
0 & \text{for } k < k_{\rm min}, \\
r \cdot A_{\rm s} \cdot 
\left(\frac{k}{k_{\rm pivot}}\right)^{n_{\rm T}} & \text{for } k \geq k_{\rm min}\;,
\end{cases}
\label{Primordial_Tensor}
\end{equation}
in terms of the amplitude, $A_{\rm s}$, of the primordial power spectrum, the spectral index 
$n_{\rm s}$ of the scalar fluctuations, the tensor-to-scalar ratio $r$ and the spectral index 
$n_{\rm T}$ of the tensor fluctuations, which we assume to be the conventional 
$n_{\rm T} = -r/8$ and the pivot scale $k_{\rm pivot}=0.05\;{\rm Mpc}^{-1}$. The value of 
$k_{\rm min}$ is given in Equation~(\ref{kmin}). The rest of the parameters are derived from 
the most recent {\it Planck} results.

For the angular correlation function, the calculation of the $E$-mode self-correlation, 
$TE$ cross-correlation, and the $B$-mode self-correlation, are performed using our own code. 
In contrast, the calculation of the temperature self-correlation, and the $Q+U$ and $Q-U$ 
modes' self-correlations, is carried out using CAMB.

\subsection{Analysis of the Temperature Signal}\label{temp}
We shall begin our analysis with the TT angular power spectrum and the corresponding
angular correlation function. These have already been published 
\citep{MeliaLopez:2018,Melia:2021b}, but our reason for reproducing these 
results is to confirm the reliability of our method before progressing
to the more important polarization features.

\begin{figure}[ht]
\centering              
\includegraphics[width=\columnwidth]{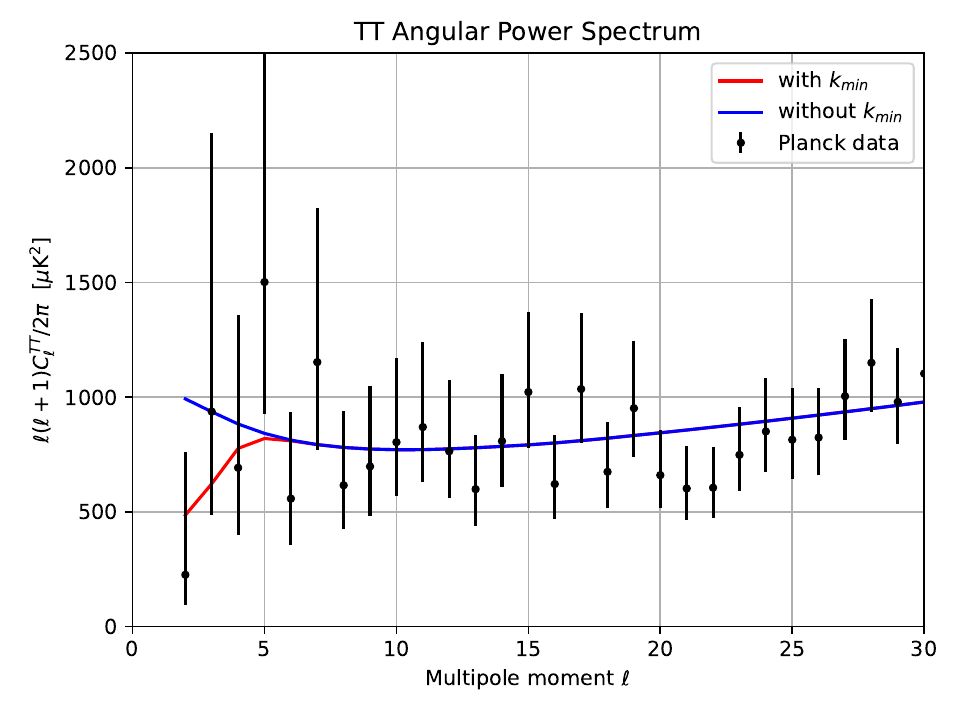}
\caption{Lensed TT angular power spectrum with tensor modes, 
for multipoles $\ell \le 30$. The blue curve includes the effects of $k_{\rm min}$, 
while the red curve represents the standard model's prediction without $k_{\rm min}$. 
The {\it Planck}-2018 data points with error bars are shown in black.}
\label{fig:CL_TT}
\end{figure}

The angular power spectrum $C_\ell^{TT}$ shown in Figure~\ref{fig:CL_TT} was 
calculated using the CAMB code with the setup we described above. We can clearly see the 
low power anomaly in this figure, especially for $\ell=2$ and $\ell=4$, where the values 
measured by {\it Planck} \citep{PlanckVI:2020}
are significantly below the values predicted by the standard model with $k_{\rm min}=0$, even 
falling outside the $1\sigma$ region of the measurement. However, once we introduce $k_{\rm min}$ 
into the analysis, the predicted values align much more closely with the measured ones. We want 
to emphasize that the two theoretical curves shown here are not fitted to the {\it Planck} data. 
They are directly calculated with CAMB using the most recent cosmological parameters optimized 
by {\it Planck} \citep{PlanckVI:2020}. Another important observation from this figure is that 
$k_{\rm min}$ does not 
affect the angular power spectrum at $\ell > 5$. In this range, the two theoretical curves are 
identical. In other words, for the TT angular power spectrum, introducing $k_{\rm min}$ helps 
explain the lack of large-angle correlation anomaly, but does not harm the part of the angular 
power spectrum that was performing very well in terms of fitting the data with the standard model.

Throughout this paper, the value of $k_{\rm min}$ we use is obtained from the 
optimization of the $TT$ angular correlation function. Therefore, in terms of statistical 
analysis, we focus on the angular correlation functions. While we include the angular power 
spectrum plots for completeness, a quantitative comparison between models is performed only 
for the angular correlation functions.

\begin{figure}[ht!]
\centering
\includegraphics[width=\columnwidth]{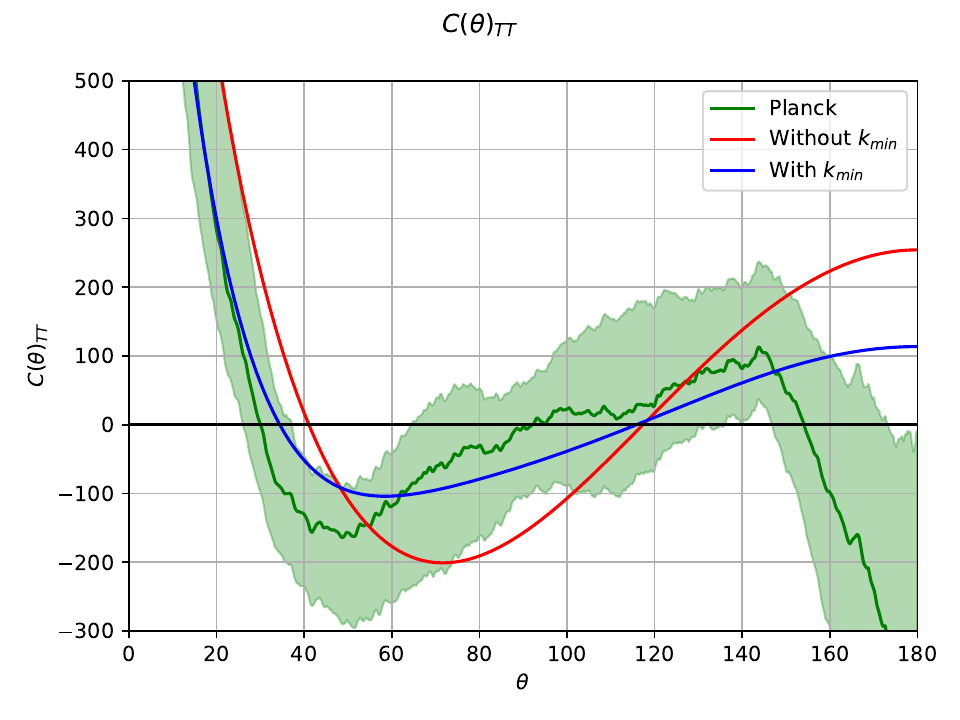}
\caption{TT angular correlation function with (blue) and without (red)
$k_{\rm min}$, in comparison with the angular correlation function
calculated from the {\it Planck} data (green). The shaded region represents
the (non-Gaussian) $1\sigma$ uncertainty obtained from a sample of three
thousand mock correlation functions generated via Monte Carlo randomization.}
\label{fig:TTCtheta}
\end{figure}

\begin{figure}[ht]
\centering
\includegraphics[width=\columnwidth]{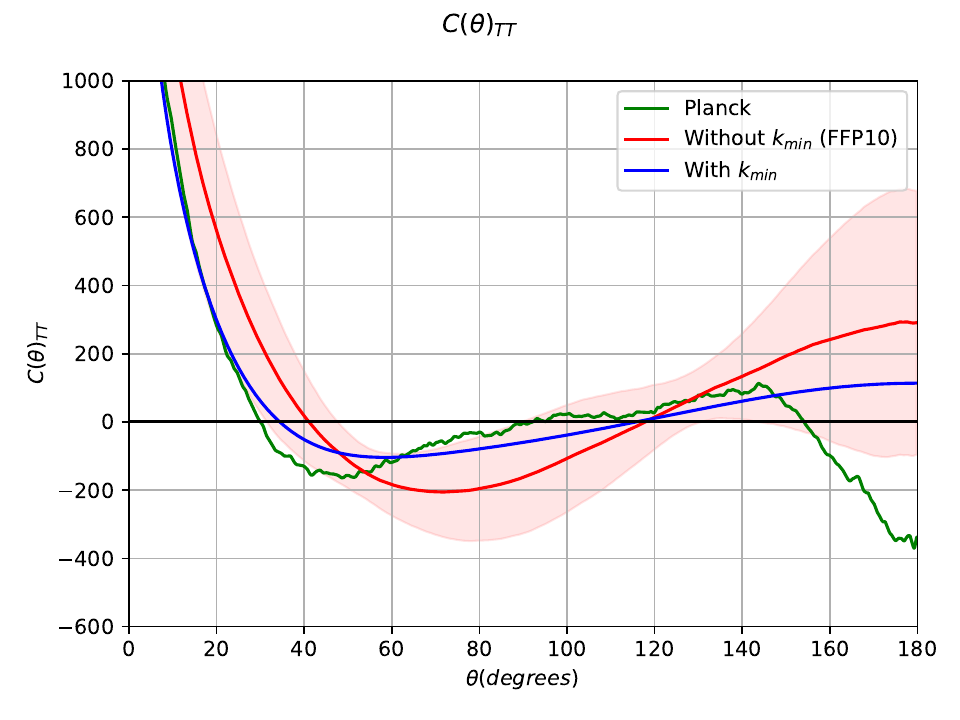}
\caption{Same as Fig.~\ref{fig:TTCtheta}, except that here the
shaded region (red) corresponds to the $1\sigma$ cosmic variance, calculated 
utilizing {\it Planck}'s official set of simulations (FFP10) for the
concordance model (red curve).}
\label{fig:TTCtheta2}
\end{figure}

\begin{figure}[ht]
\centering
\includegraphics[width=\columnwidth]{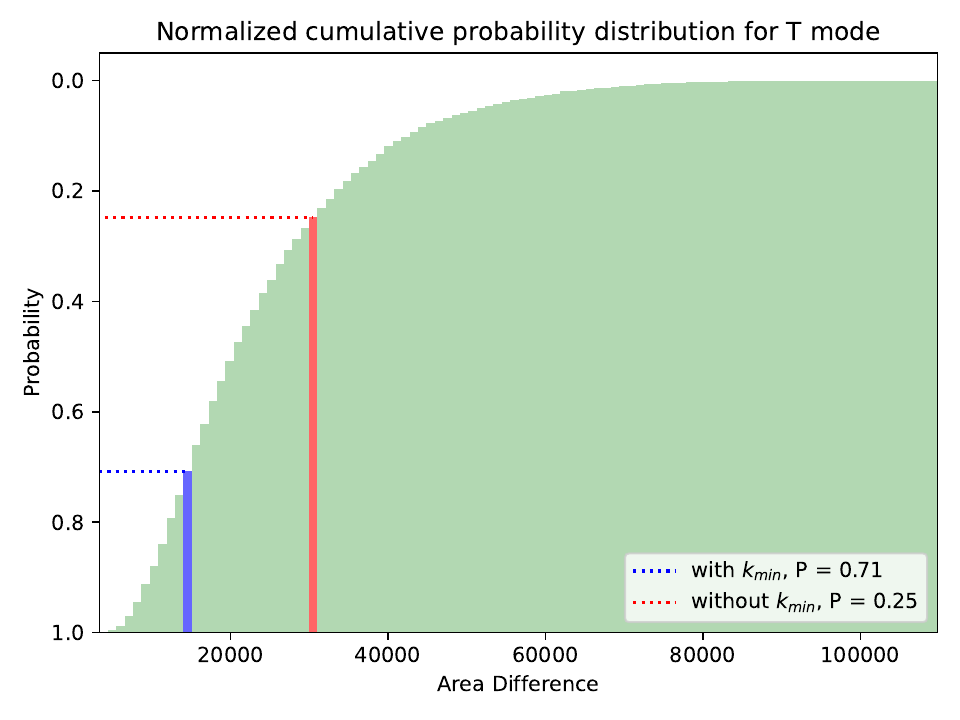}
\caption{The cumulative probability distribution of the temperature mode, 
generated from the area differences between the {\it Planck} curve and the three thousand 
realizations produced using {\it Planck}'s total errors. Blue represents the model with 
$k_{\rm min}$; red represents the model without $k_{\rm min}$.}
\label{fig:accuPTT}
\end{figure}

The TT angular correlation function is shown in Figures~\ref{fig:TTCtheta} and 
\ref{fig:TTCtheta2}. In Figure~\ref{fig:TTCtheta}, we display the two theoretical 
curves with and without $k_{\rm min}$, calculated with CAMB using the angular power 
spectrum in Figure~\ref{fig:CL_TT}. The {\it Planck} curve was also calculated with CAMB 
using the power spectrum data in Figure~\ref{fig:CL_TT}. In addition, 
Figure~\ref{fig:TTCtheta} includes the $1\sigma$ uncertainty associated 
with the {\it Planck} curve, achieved by generating three thousand mock 
realizations of the angular correlation function. Specifically, for each $\ell$, we used 
the measured value of $C_{\ell}^{TT}$ and its corresponding errors to generate a mock 
sample of this $C_{\ell}^{TT}$. To address the non-Gaussianity of the data, we assumed 
that the upper and lower errors of $C_{\ell}^{TT}$ satisfy two different half-Gaussian 
distributions, and then randomized the $C_{\ell}^{TT}$ values within these distributions. 
This process yielded a complete mock set of $C_{\ell}^{TT}$, from which we calculated a 
mock $C(\theta)$ curve. We repeated this process three thousand times to 
generate three thousand mock $C(\theta)$ curves, and then determined the $1\sigma$ 
range of $C(\theta)$ at every 
angle using these samples. To properly address the non-Gaussianity (because these 
$C(\theta)$ samples were generated from non-Gaussian distributions, they obviously carry 
non-Gaussianity), we assumed that the upper and lower errors of $C(\theta)$ at every 
angle satisfy two different half-Gaussian distributions. In other words, the $1\sigma$ 
region was calculated separately for the upper and lower errors.

From Figure~\ref{fig:TTCtheta}, we can clearly see the lack of large-angle correlations
in the data. The theoretical curve without $k_{\rm min}$ misses the {\it Planck} curve at 
all angles and even falls outside of the $1\sigma$ error region over some of the plot.
This situation is notably worse than the deficit of power at low $\ell$'s seen in the
angular power spectrum. This is because (as noted earlier) every $C_\ell$ contributes 
at all angles. Thus, the discrepancy between measured and theoretical $C_\ell$ values 
(which only appears at low $\ell$'s) is observed at all angles in the angular correlation 
function. After introducing $k_{\rm min}$, we see that the theoretical curve aligns much 
more consistently with the measured curve---at all angles. The cutoff $k_{\rm min}$ not 
only improves the performance of the theoretical curve at angles $> 60$ degrees, but 
actually also enhances the fit at $< 60$ degrees. After introducing $k_{\rm min}$, the 
theoretical curve falls within the $1\sigma$ error region throughout the plot.

The TT angular correlation function is shown again in Figure~\ref{fig:TTCtheta2}, 
but this time with an alternative method of analysis. This time, the $1 \sigma$ region 
corresponds to the theoretical curve without $k_{\rm min}$, based solely on cosmic 
variance, estimated using the {\it Planck} official simulations (FFP10 CMB realizations). 
Specifically, we took a thousand simulated CMB maps from the FFP10 \citep{PlanckVI:2020} 
simulation, used Healpy \citep{Gorski:2005} to read these maps, and calculated their 
angular power spectrum. We then used 
CAMB to calculate their angular correlation function. The samples we used only included 
the cosmic variance as the source of error, which is assumed to be Gaussian. The 
theoretical curve without $k_{\rm min}$ and the corresponding $1 \sigma$ region were 
then calculated, respectively, by taking the average and the uncertainty of these one 
thousand samples at each angle. The theoretical curve with $k_{\rm min}$ is the same 
as the one shown in Figure~\ref{fig:TTCtheta}. Figure~\ref{fig:TTCtheta2} 
also shows that introducing a non-zero $k_{\rm min}$ significantly improves the 
consistency between the observations and theoretical predictions.

This visual comparison showing that the introduction of $k_{\rm min}$ 
significantly improves the alignment between the model and the measurements may
be made more quantitative, as described below. Notably, we shall compare three curves: 
two generated from theoretical predictions and one derived from measurements. Statistical
tests based on how well one curve matches another tend to be based on area differences.
We shall adopt a technique introduced for such situations in \cite{Melia:2018g}. 

Specifically, we calculate the area difference between the {\it Planck} curve and the 
3,000 realizations used to estimate its $1 \sigma$ region. We then compute and plot 
the cumulative probability distribution, shown in Figure~\ref{fig:accuPTT}. This 
figure actually represents the cumulative frequency of the three thousand realizations 
versus the area difference, divided into 100 bins. It allows us to estimate the 
probability that a given area difference arises from the total errors of the 
measurements, which were used to generate the three thousand realizations.

We then calculate the area differences between the two theoretical curves and the 
{\it Planck} curve. Next, we identify the corresponding bins on the cumulative 
probability distribution plot, and estimate the probability that the area 
difference between a theoretical curve and the {\it Planck} curve arises solely 
from the total error of the {\it Planck} measurements. The results show that, for the 
theoretical curve without $k_{\rm min}$, the probability that the area difference is 
due to the total errors of the {\it Planck} measurements is 0.25. For the curve with 
$k_{\rm min}$, the probability is 0.71. These results clearly indicate that introducing 
$k_{\rm min}$ significantly reduces the discrepancy between theory and observation.

The analysis and results we have presented here confirm those obtained by the previously
published work which, however, used a different method of analysis and a different set of data 
({\it Planck}-2014). Introducing $k_{\rm min}$ has significantly brought the theoretical 
predictions closer to the measurements at the low $\ell$ end of the angular power spectrum 
and at all angles of the angular correlation function.

\begin{figure}[ht!]
\centering
\includegraphics[width=\columnwidth]{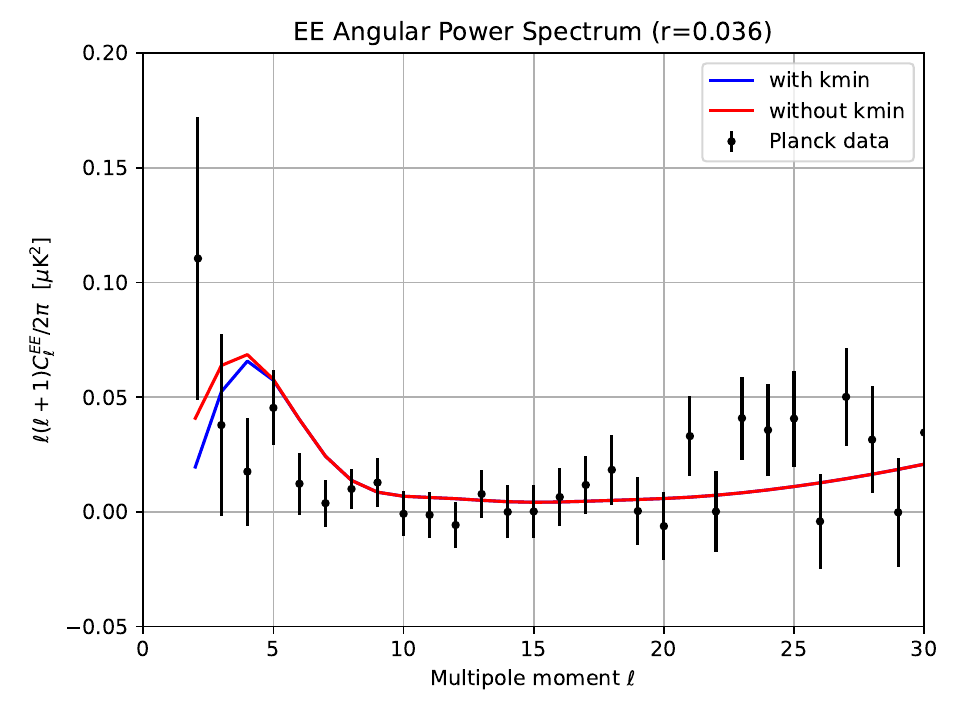}
\includegraphics[width=\columnwidth]{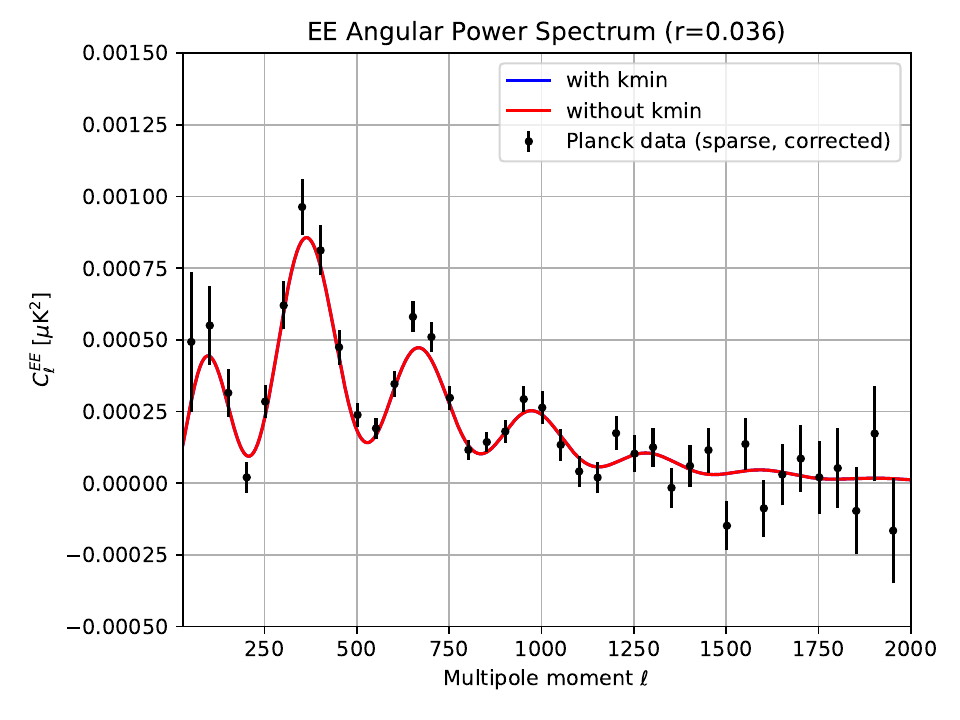}
\caption{Lensed power spectra with tensor modes. Top: the $EE$ power spectrum 
for multipoles $\ell \leq 30$. Bottom: the $EE$ power spectrum for multipoles 
$\ell > 30$. In both cases, the blue curve includes the effects of $k_{\rm min}$, 
while the red curve represents calculations without it, i.e., the conventional
application of the standard model. The {\it Planck}-2018 data points with error 
bars are shown in black.}\label{fig:CL_EE}
\end{figure}

\subsection{Analysis of the E-Mode Polarization}\label{E-Mode}
Let us now proceed to an analysis of the $E$-mode polarization. Again, we calculate 
the $E$-mode angular power spectrum and the angular correlation function. It is important 
to note that when calculating the $E$-mode polarization, both the scalar and tensor modes 
contribute to the final results, which are also impacted by lensing effects. We use CAMB 
and {\it Planck}'s optimized parameters \citep{PlanckVI:2020} to calculate these 
E-mode quantities. For the spectral index $n_T$ of the tensor fluctuations, we employ 
the conventional assumption that $n_T = -r/8$. For the tensor-to-scalar ratio $r$, we 
initially compare the outcomes for two different values, $r=0.00461$ and $r=0.036$. 
We choose $r=0.036$ because it represents the current upper limit from the analysis 
of the {\it Planck} and BICEP2/Keck data \citep{Tristram:2021}. We choose $r = 0.00461$ 
to be consistent with the $B$-mode analysis that we shall motivate later. As it turns out, 
the angular power spectra resulting from this procedure are extremely similar in spite 
of the different $r$ values, presumably because the $E$-modes are predominantly sourced 
by scalar rather than tensor perturbations. For the following analysis, we therefore 
retain only the results corresponding to $r=0.036$.

\begin{figure}
\centering    
\includegraphics[width=\columnwidth,height=5.6cm]{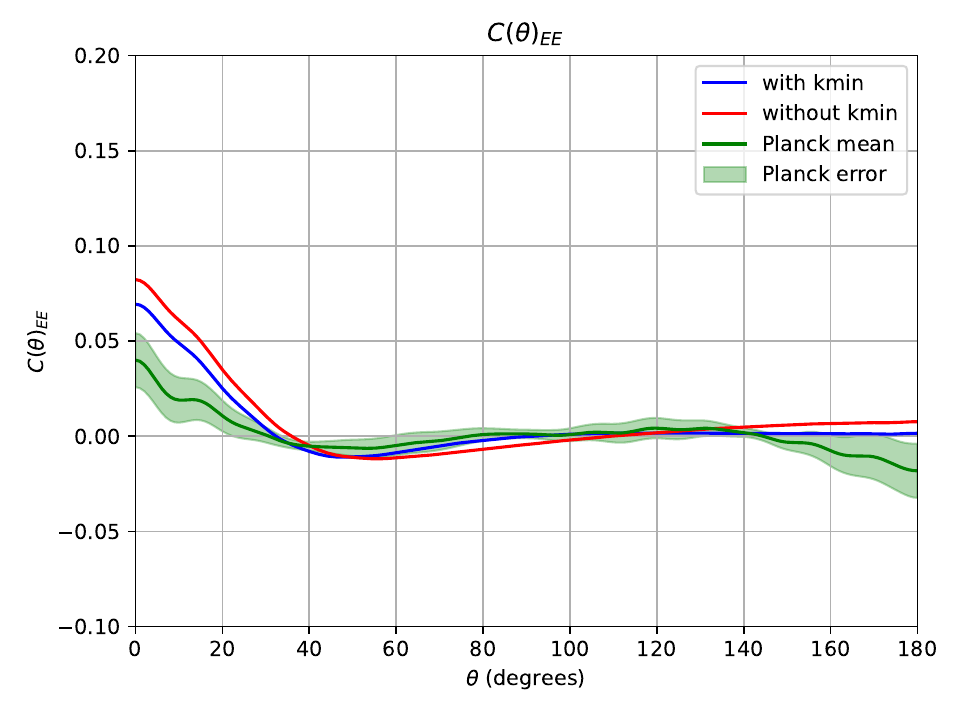}
\includegraphics[width=\columnwidth,height=5.6cm]{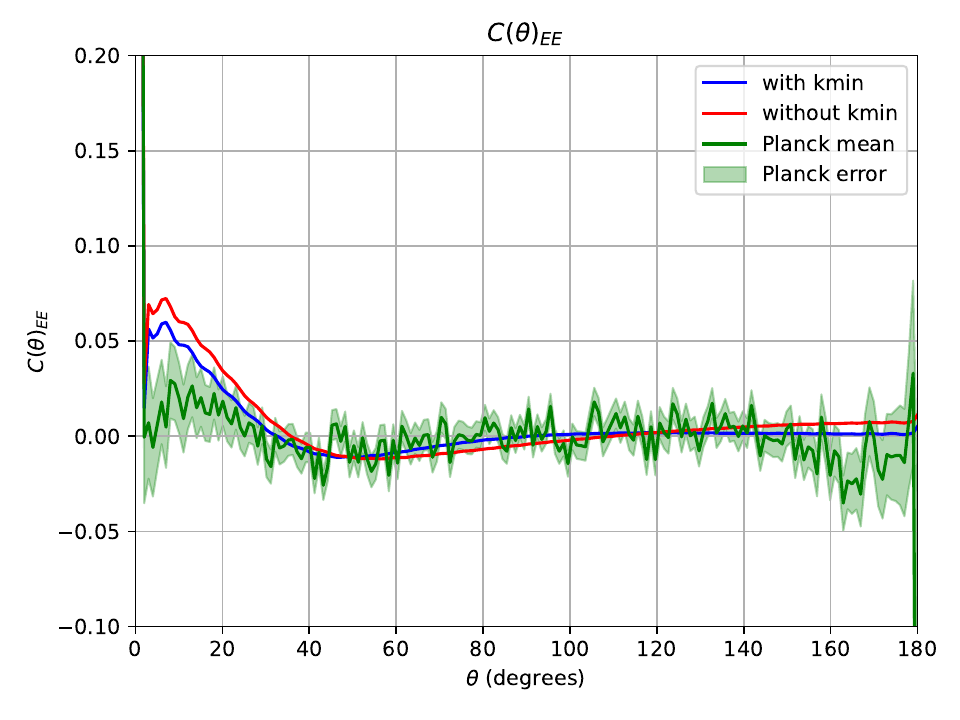}
\includegraphics[width=\columnwidth,height=5.6cm]{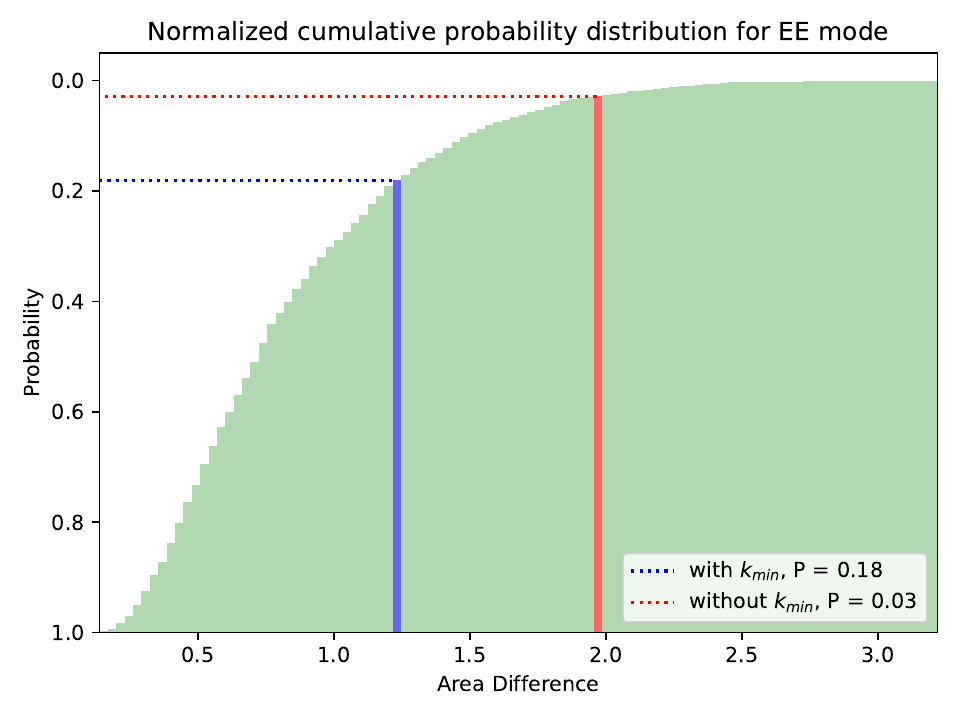}
\caption{$EE$ angular correlation function. Top: the $EE$ angular 
correlation functions including $\ell \leq 30$ modes. Middle: the $EE$ angular 
correlation functions including $\ell \leq 1500$ modes. In both cases, the blue 
curve includes the effects of $k_{\text{min}}$, while the red curve represents 
the conventional application of the standard model without $k_{\text{min}}$. 
The green curve represents the angular correlation function 
calculated from the {\it Planck}-2018 data. The corresponding  $1\sigma$ region 
was generated from a three thousand step MC simulation utilizing the total errors in 
{\it Planck}'s measurements. Bottom: The cumulative probability distribution of the 
EE mode (including $\ell \leq 30$ modes only), generated from the area differences 
between the {\it Planck} curve and the three thousand realizations mentioned above, 
which were produced using {\it Planck}'s total errors. Blue represents the model with 
$k_{\rm min}$; red represents the model without $k_{\rm min}$.}\label{fig:Ctheta_EE}
\end{figure}

The angular power spectrum $C_{\ell}^{EE}$ is shown in Figure~\ref{fig:CL_EE}. One can easily confirm that (i) the low power anomaly 
not only appears in the temperature angular power spectrum, but is also evident here, 
for $\ell < 7$; (ii) the theoretical prediction is always greater than the measured 
values, except for $\ell=2$. It is worth noting, however, that {\it Planck}'s 
MC simulations are not able to quantify the residual systematics at $\ell = 2$ 
\citep{PlanckVII:2020}; (iii) similarly to the situation with the temperature 
angular power spectrum, $k_{\rm min}$ helps to lower the power at low $\ell$'s. The 
improvement, however, is not as significant as in the temperature case. After introducing 
$k_{\rm min}$, the predicted values are still somewhat higher than the measured ones; 
and (iv) another important result of this calculation is that the two theoretical 
curves (with and without $k_{\rm min}$) are extremely similar at $\ell > 5$, indicating 
that, while $k_{\rm min}$ helps resolve the low power anomaly at low $\ell$'s, it does 
not affect the success of the conventional models at high $\ell$'s.

Calculation of the angular correlation function from the angular power spectrum is 
straightforward, given that \citep{Baumann:2009} 
\begin{equation}
\begin{aligned}
C_{EE}(\theta) &= \sum_{\ell} \frac{2\ell + 1}{4\pi} C_{\ell}^{EE} P_{\ell}(\cos \theta)\,, \\
C_{BB}(\theta) &= \sum_{\ell} \frac{2\ell + 1}{4\pi} C_{\ell}^{BB} P_{\ell}(\cos \theta)\,.
\end{aligned}
\label{CthetaEB}
\end{equation}
The results are shown in Figure~\ref{fig:Ctheta_EE}, which shows the EE self-angular 
correlation function calculated from the angular power spectrum in Figure~\ref{fig:CL_EE}. 
The results of models with and without $k_{\rm min}$ correspond to blue and red, 
respectively. The {\it Planck} measurements are plotted in green, along with their 
$1\sigma$ error regions, calculated from {\it Planck}'s published $EE$ angular power 
spectrum data \citep{PlanckVI:2020} through a three thousand step Monte Carlo (MC) 
simulation. To be more specific, we took the observed values of the $EE$ angular power 
spectrum and their errors (which were published as Gaussian errors), and then generated 
three thousand sets of mock $C_{\ell}^{EE}$'s. An angular correlation function was 
calculated for each set, and from these we then determined the average and standard 
deviation of the angular correlation function at each angle. 

We performed this analysis 
for the range $\ell \leq 30$ because, as observed from the angular power spectrum of 
the $E$-mode polarization, the impact of $k_{\rm min}$ is significant only at $\ell < 
6$. Additionally, including higher $\ell$ terms makes the angular correlation function 
fuzzy because high-$\ell$ Legendre polynomials oscillate significantly with respect to 
angle. Therefore, we believe that including only a limited range of $\ell$'s is 
beneficial for demonstrating the impact of $k_{\rm min}$ on the angular correlation 
function. We also present the results for $\ell \leq 1500$ here (middle panel of 
Figure~\ref{fig:Ctheta_EE}) to make our results more complete and to support the above
conclusion. As we can see from the plot, the trend of the three curves is very 
similar to the $\ell \leq 30$ case. However, the {\it Planck} curve appears fuzzy and 
appears to be unstable at very small and large angles (close to 0 and 180 degrees). 
Therefore, the quantitative analysis will be performed on the $\ell \leq 30$ case. For 
the rest of this paper, we shall always limit the range to $\ell \leq 30$ for the analysis 
of the angular correlation function associated with the $E$ mode.

As we can see from the $EE$ angular correlation plots, the conventional model without 
$k_{\rm min}$ does not fit the observed data at all. Similarly to the temperature case, 
it is too large at $\theta < 40^\circ$, too small at $60^\circ < \theta < 120^\circ$, 
and too large again at $\theta > 140^\circ$. On the $\ell < 30$ plot, it even 
misses the $1\sigma$ error region at most angles. After introducing $k_{\rm min}$, the 
theoretical curve is significantly closer to the observed one. Although it is still too 
large at $\theta < 30^\circ$, it fits the observed curve very well at $60^\circ < \theta 
< 120^\circ$; more importantly, it lies within the $1\sigma$ error region of the observed 
curve at most angles.

Similar to the case of the $TT$ angular correlation functions, we are 
comparing three curves: two from theoretical predictions and one calculated from 
measurements. To quantitatively assess differences between these curves, we again use a
cumulative probability approach to estimate the likelihood that the area difference 
between a theoretical curve and the {\it Planck} curve arises solely from the total 
errors of the {\it Planck} measurements. The results are shown in the bottom
panel of Figure~\ref{fig:Ctheta_EE}. For the theoretical curve without $k_{\rm min}$, 
the probability that the area difference is due to the total errors of {\it Planck} 
is 0.03. For the curve with $k_{\rm min}$, the probability is a much improved 0.18.

It is very clear from the analysis of the $EE$ angular power spectrum and angular correlation 
function that we confirm the results of previous work based on the CMB temperature. Introducing 
$k_{\rm min}$ significantly helps in bringing the theoretical predictions closer to the observed 
results. It is worth noting, however, that although the analysis of the $E$-mode makes the 
conclusion more robust with the inclusion of polarization data, the $E$-mode is still dominated 
by scalar perturbations. As mentioned earlier, if $k_{\rm min}$ truly exists and is due to the 
onset of slow-roll inflation, it should impact not only the scalar but tensor modes as well. 
We shall thus analyze the $B$-mode polarization shortly.

\subsection{Analysis of the TE Cross Correlation}\label{TE}
We next consider the $TE$ cross-correlation between the temperature and $E$-mode polarization 
in the CMB. This cross-correlation provides substantial information about the reionization 
era. Analyzing the $TE$ mode is essential for breaking the 
degeneracies between various cosmological parameters, thereby leading to more precise 
measurements. Furthermore, the $TE$ mode serves as an instrumental tool for testing 
cosmological models, as it offers an independent check of the results obtained from 
the temperature and $E$-mode polarization power spectra separately.

Again, we will begin our analysis with the angular power spectrum of the $TE$ mode, and 
then move on to the angular correlation function. When calculating the $TE$ cross-correlation, 
both scalar and tensor modes contribute to the final result, and is also influenced by the 
lensing effect. We use CAMB along with {\it Planck}'s optimized parameters 
\citep{PlanckVI:2020} to calculate the angular power spectrum. Similarly to the $E$-mode 
case, we adopt the conventional assumption that $n_T = -{r}/{8}$. The tensor-to-scalar 
ratio $r = 0.036$ is chosen to be consistent with the analysis focusing solely on $E$-mode.

\begin{figure}
\centering
\includegraphics[width=\columnwidth]{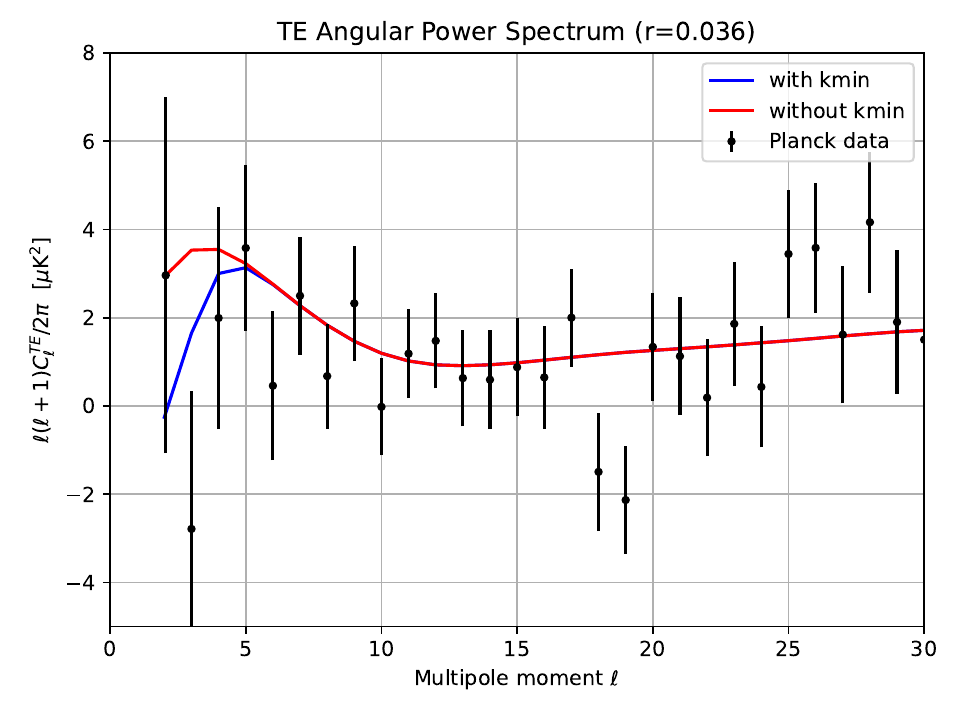}
\includegraphics[width=\columnwidth]{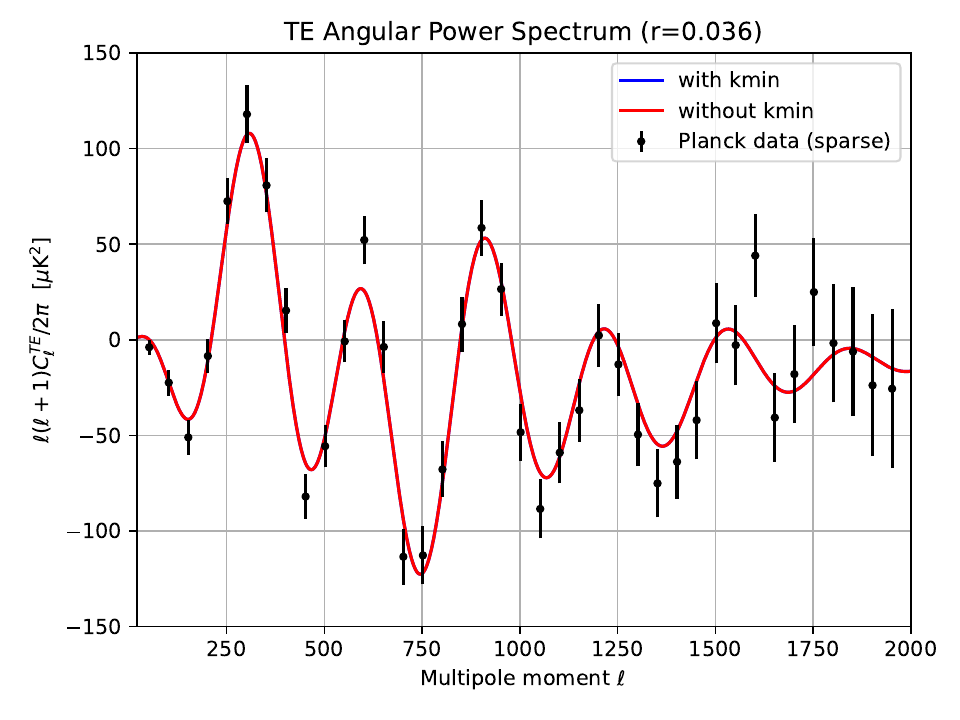}
\caption{$TE$ Lensed power spectra with tensor modes. Top: the $TE$ power spectrum 
for multipoles $\ell \leq 30$. Bottom: the $TE$ power spectrum for multipoles 
$\ell > 30$. In both cases, the blue curve includes the effects of a minimum wavenumber 
$k_{\rm min}$, while the red curve represents calculations without it, i.e., the
conventional application of the standard model. The {\it Planck}-2018 data 
points with error bars are shown in black.}\label{fig:CL_TE}
\end{figure}

\begin{figure}[ht!]
\centering    
\includegraphics[width=\columnwidth]{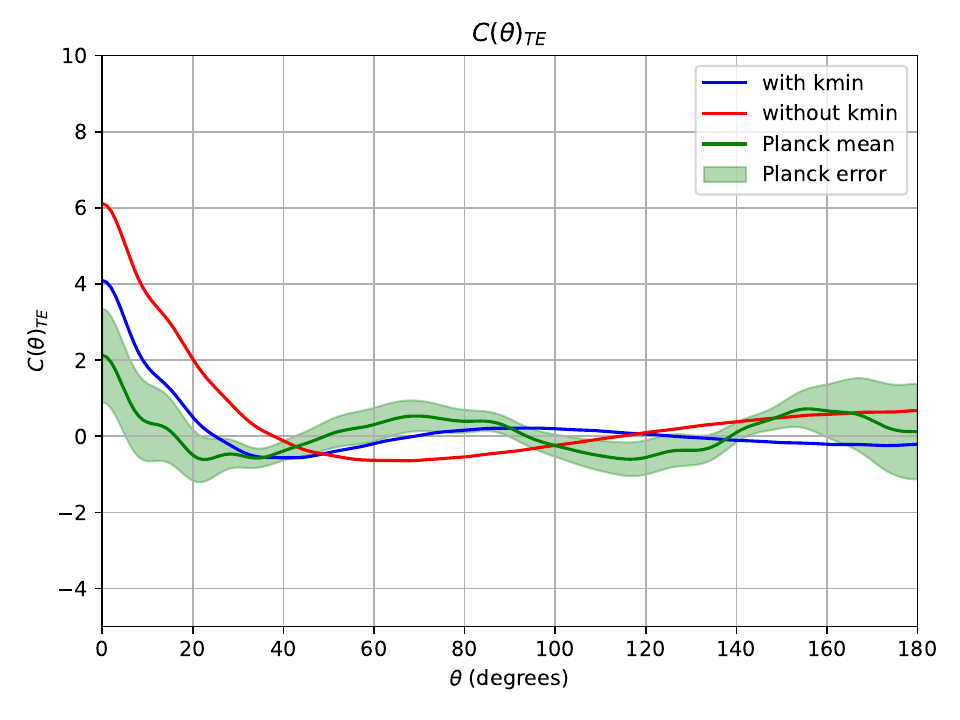}
\includegraphics[width=\columnwidth]{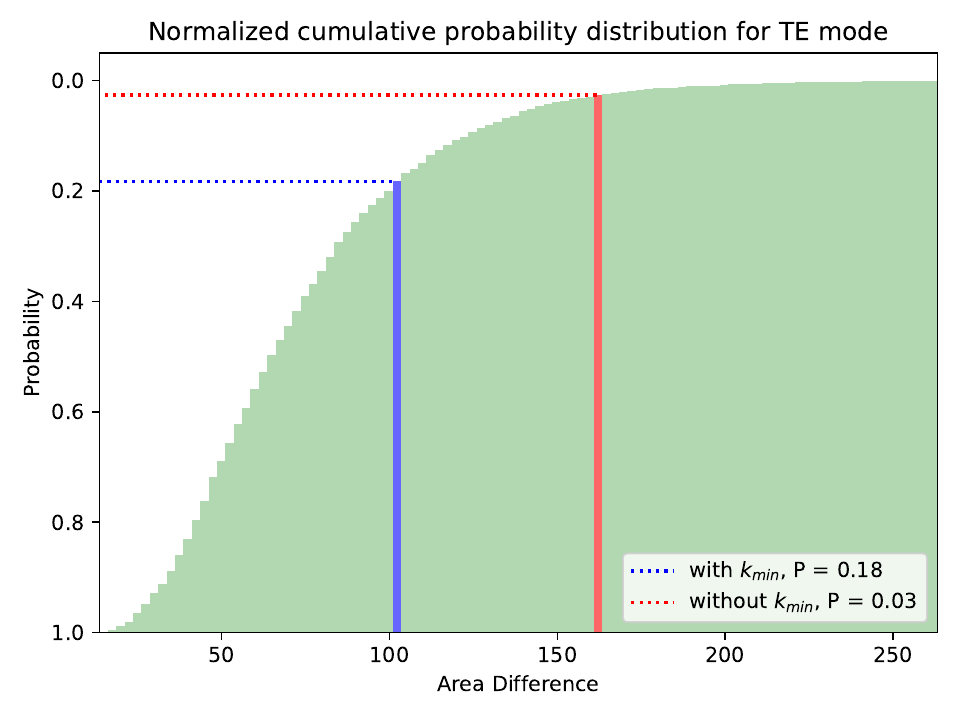}
\caption{$TE$ angular correlation function. Top: the $TE$ angular correlation 
function including $\ell \leq 30$ modes. The blue curve includes the effects of $k_{\text{min}}$, 
while the red curve represents calculations without $k_{\text{min}}$, i.e., the conventional
application of the standard model. The green curve represents the angular correlation
function calculated from the {\it Planck}-2018 data. The corresponding  $1\sigma$ region 
was generated from a three thousand step MC simulation utilizing the total errors in 
{\it Planck}'s measurements. Bottom: The cumulative probability distribution of the $TE$ 
mode, generated from the area differences between the {\it Planck} curve and the three 
thousand realizations mentioned above, which were produced using {\it Planck}'s total 
errors. Blue represents the model with $k_{\rm min}$, red represents the model without 
$k_{\rm min}$.}\label{fig:Ctheta_TE}
\end{figure}

The results of the angular power spectrum are shown in Figure~\ref{fig:CL_TE}. 
The top panel displays the low $\ell$ region ($\ell \leq 30$), while the bottom panel shows 
the rest ($30 < \ell < 2000$). In both plots, the blue curve includes the effects of a 
minimum wavenumber $k_{\rm min}$, while the red curve represents calculations without 
$k_{\text{min}}$. Additionally, the most updated {\it Planck} data are presented 
along with their errors in black. 

The first observation one can make from an inspection of these plots is that the low 
power anomaly at low $\ell$'s in the angular power spectrum is very apparent. Though 
the error bars for the first few $\ell$'s are quite large, the downward trend in the
power spectrum is very pronounced, and is much steeper than the theoretical curve
without $k_{\rm min}$. The $\ell = 2$ mode appears to be an exception; 
however, as mentioned earlier, {\it Planck}'s MC simulations cannot quantify the 
residual systematics at $\ell = 2$.

A second observation is that, similarly to the temperature case, there is an obvious 
difference at $\ell < 6$ between the theoretical curves with and without $k_{\rm min}$. 
The curve with the cutoff is obviously steeper and fits the data much better, though at 
$\ell=3$, the prediction still falls outside the $1 \sigma$ error region. This feature 
in $C_{\ell}^{TE}$ is more pronounced than in $C_{\ell}^{EE}$. Third, there is almost 
no difference between the two theoretical curves at $\ell>6$, as was the case for the 
temperature and $E$-mode polarization.

The equation used to calculate the $TE$ cross-correlation function is \citep{Baumann:2009} 
\begin{equation}
\begin{aligned}
C_{TE}(\theta) &= \sum_{\ell} \frac{2\ell + 1}{4\pi} C_{\ell}^{TE} P_{\ell}(\cos \theta) \;.
\end{aligned}
\label{Ctheta_TE}
\end{equation}
To ensure consistency with the $E$-mode analysis, we select the range of $\ell$ 
for our calculations to be $\ell \leq 30$. The results for the models, both with and
without $k_{\rm min}$, are obtained using Equation~(\ref{Ctheta_TE}) with the $TE$ angular 
power spectrum computed via CAMB, based on the latest {\it Planck} optimized parameters 
\citep{PlanckVI:2020}, 
and are plotted in Figure~\ref{fig:CL_TE}. The {\it Planck} curve and its $1\sigma$ 
confidence region are derived from {\it Planck}'s published $TE$ angular power 
spectrum data through a three-thousand step Monte Carlo (MC) simulation. We utilize the 
observed values of the $TE$ angular power spectrum and their reported Gaussian errors 
to generate three thousand mock sets of $C_{\ell}^{TE}$. An angular correlation 
function is then calculated for each mock set, and we compute the mean and standard deviation 
of the angular correlation functions at each angle.

Our results are shown in Figure~\ref{fig:Ctheta_TE}. The theoretical prediction 
without $k_{\rm min}$ is displayed in red, while the case with $k_{\rm min}$ is shown 
in blue. The curve calculated from {\it Planck}'s measurements is presented in 
green, along with its $1\sigma$ error region. Unlike the temperature case,
there is no lack of large-angle correlation, but the theoretical prediction without 
$k_{\rm min}$ still deviates significantly from the measurements at almost all angles, 
crossing into the $1\sigma$ region in only a few limited areas. After introducing 
$k_{\rm min}$, however, the theoretical fit improves significantly at most angles, 
bringing it much closer to the measurements.

Again, we quantitatively compare the fits using a cumulative probability 
approach to estimate the probability that the area difference between a theoretical 
curve and the {\it Planck} curve arises solely from the total errors in {\it Planck}. 
The results are shown in the bottom panel of Figure~\ref{fig:Ctheta_TE}. For the 
theoretical curve without $k_{\rm min}$, the probability that the area difference 
is due to the total errors of the {\it Planck} measurements is 0.03. For the curve 
with $k_{\rm min}$, the probability is a much improved 0.18.

The results obtained from the $TE$ angular power spectrum and cross-correlation function 
clearly confirm the conclusions drawn by the previous study based solely on the temperature,
and also with the earlier analysis in this paper, i.e., the low power anomaly is 
quite pronounced in the $TE$ angular power spectrum. Introducing $k_{\rm min}$ to the 
conventional model significantly helps to align the theoretically predicted curve with
the observations, and has no impact on the predictions at $\ell>6$, where the conventional 
model is already very successful. The difference between the cases with and without 
$k_{\rm min}$ is definitely detectable, and the data clearly favor the model with 
$k_{\rm min}$. And though there is no lack of large-angle correlation in the $TE$ 
cross-correlation plots, the theoretically predicted $TE$ angular correlation function 
deviates significantly from the measurements, while introducing $k_{\rm min}$ helps to 
ease the tension considerably.

\subsection{Analysis of the B-mode Polarization}\label{B-mode}
Parts of the B-mode results we present here have been included in one of our previous papers 
\citep{Liu:2024PLB}. However, to make the discussion here self-contained, we shall present
the complete analysis.

A study of the $B$-mode polarization is quite different from the $E$-mode and the 
cross-correlations. On the one hand, because $B$-mode can only be generated from tensor 
modes or through the lensing effect \citep{Zaldarriaga:1997,Zaldarriaga:1998,Lewis:2006}, 
which transforms part of the $E$-mode polarization into $B$-mode, it is very sensitive 
to the tensor-to-scalar ratio $r$. Unfortunately, there is no definitive way to constrain 
this parameter very well; we only have an estimated upper limit, which comes from the 
cross-analysis of {\it Planck} and BICEP2/KECK data \citep{Tristram:2021}, giving us the 
value $r < 0.036$. Therefore, in our analysis, we will try different values of $r$. 
As we shall show later, the larger the value of $r$, the more pronounced is the 
difference we can observe between the cases with and without $k_{\rm min}$. On the 
other hand, there is currently no solid observational data for $B$-mode, so the 
theoretical predictions have no firm constraints at the moment. Thus, we shall use the 
predicted accuracy of the upcoming LiteBIRD \citep{LiteBIRD:2022} mission as a reference 
to determine under what circumstances the differences between the cases with and without 
$k_{\rm min}$ are measurable with the next generation of CMB detectors.

\begin{figure}
\centering
\includegraphics[width=\columnwidth]{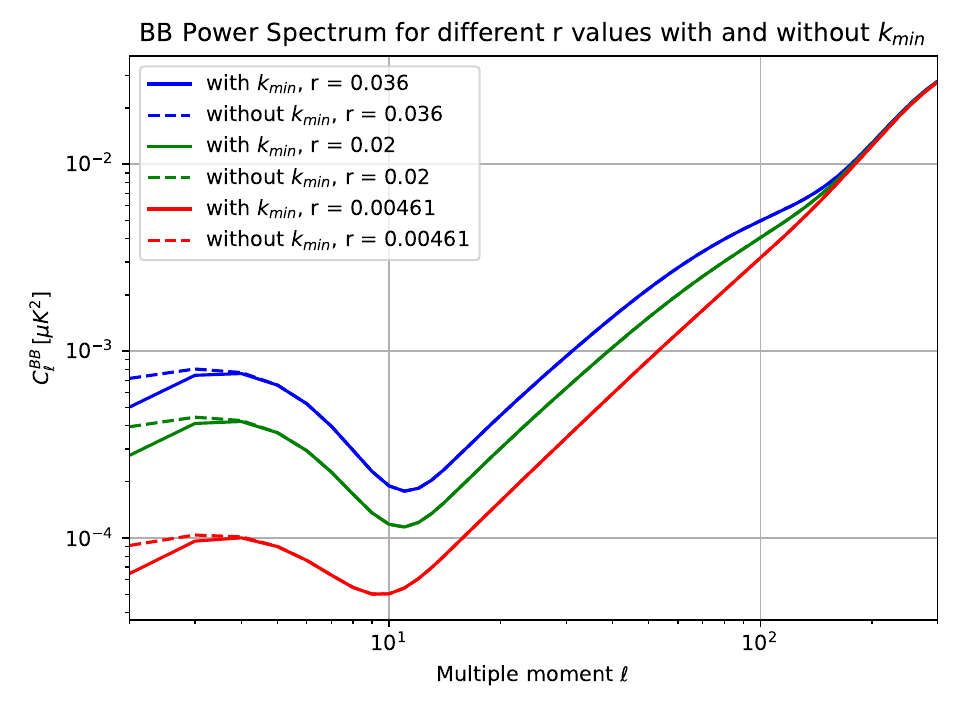}
\caption{BB angular power spectrum for different values of $r$: 
$0.00461$ , $0.02$ and $0.036$ , with (solid) and without (dashed) $k_{\rm min}$.} 
\label{fig:CL_BBvaryr}
\end{figure}

Figure~\ref{fig:CL_BBvaryr} shows how the angular power spectrum of the $B$-mode polarization
is affected by the tensor-to-scalar ratio $r$. In this figure, we have plotted both cases with 
and without $k_{\rm min}$, for three different values of $r$. These values are chosen for the 
following reasons: $r=0.036$ is the current upper limit which, as we shall demonstrate, offers 
the largest opportunity to detect the difference between the cases with and without 
$k_{\rm min}$; $r=0.00461$ is the value used by the LiteBIRD team when presenting their 
predictions of the observatory's accuracy; and as we shall see, $r=0.02$ is the value
for which the difference in angular power spectra first becomes detectable. It is quite 
evident that the larger the tensor-to-scalar ratio, the larger is the angular power spectrum 
at $\ell < 200$. Noting that this is a log-log plot, it is also very clear that the larger 
the tensor-to-scalar ratio, the more pronounced is the difference between the cases with 
and without $k_{\rm min}$.

The manner in which the tensor-to-scalar ratio $r$ affects the $B$-mode angular power spectrum 
is linked to how $B$-mode polarization is generated. As mentioned earlier, $B$-mode arises 
from tensor fluctuations and the lensing effect. While tensor modes primarily induce $B$-mode 
at large scales, the lensing effect mainly induces $B$-mode at small scales because gravitational 
lensing occurs locally in regions that are tiny compared to the entire sky. As a result, 
tensor modes mainly contribute to the low $\ell$ end of the spectrum, while the lensing effect 
contributes to the high $\ell$ end.

\begin{figure}[ht!]
\centering    
\includegraphics[width=0.99\columnwidth]{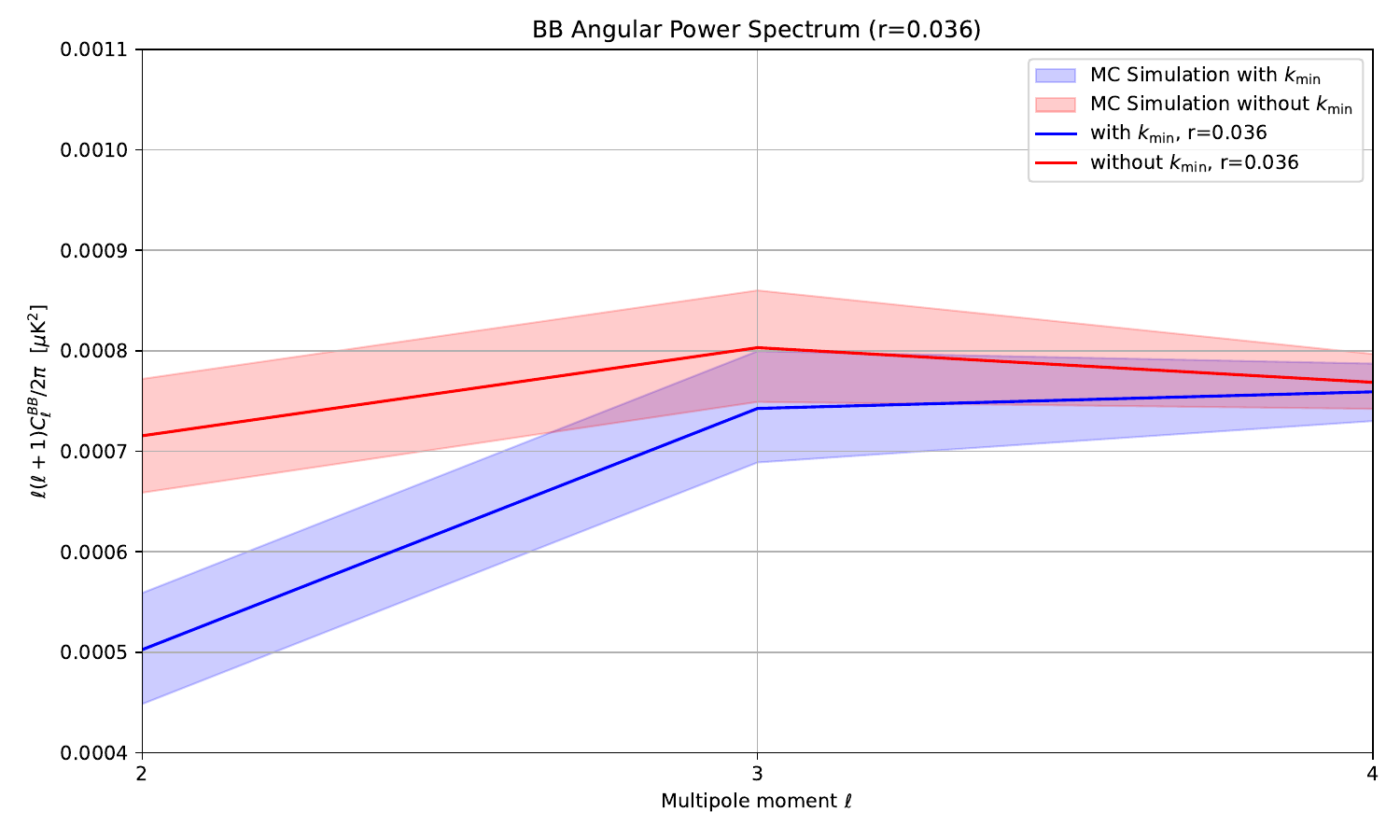}
\includegraphics[width=0.99\columnwidth]{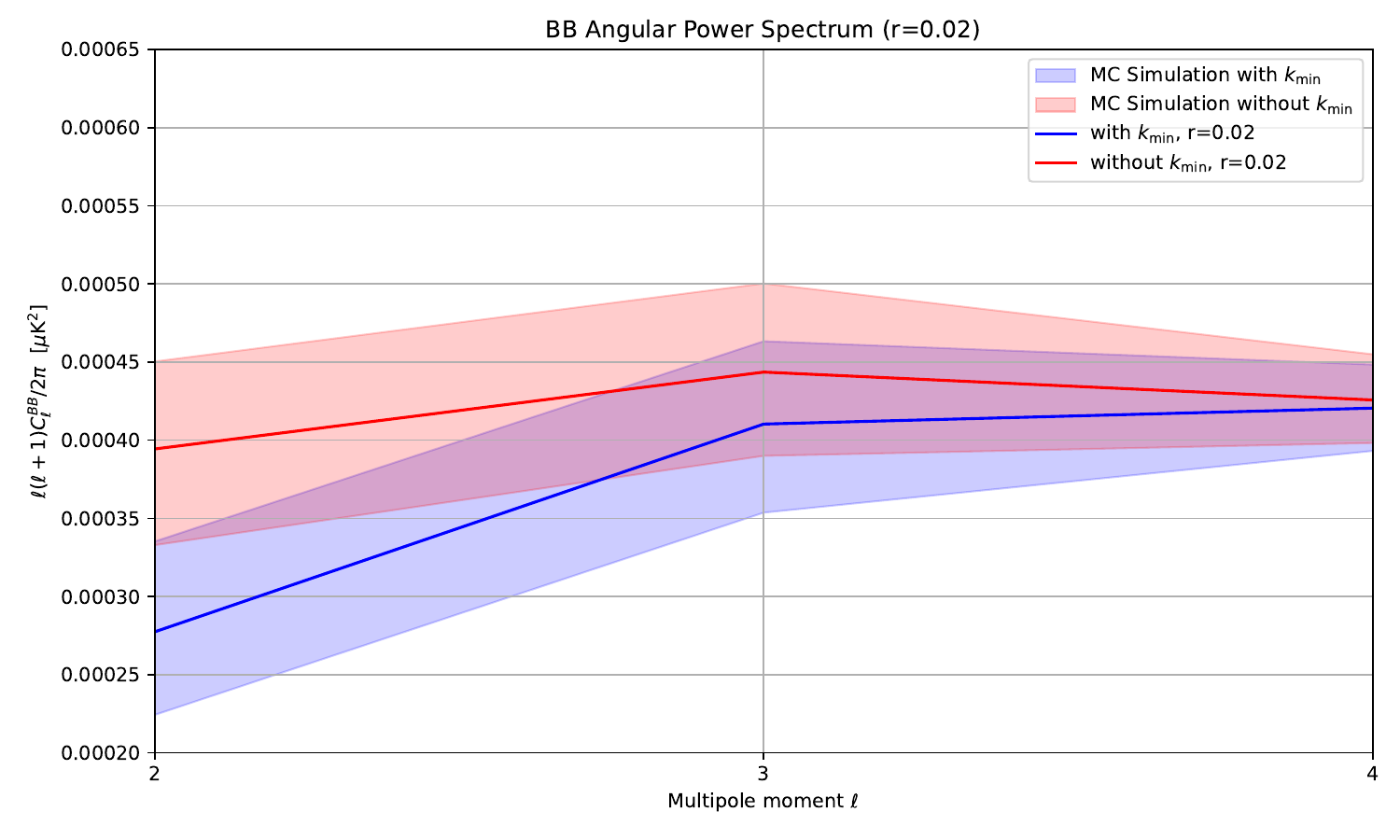}
\includegraphics[width=0.99\columnwidth]{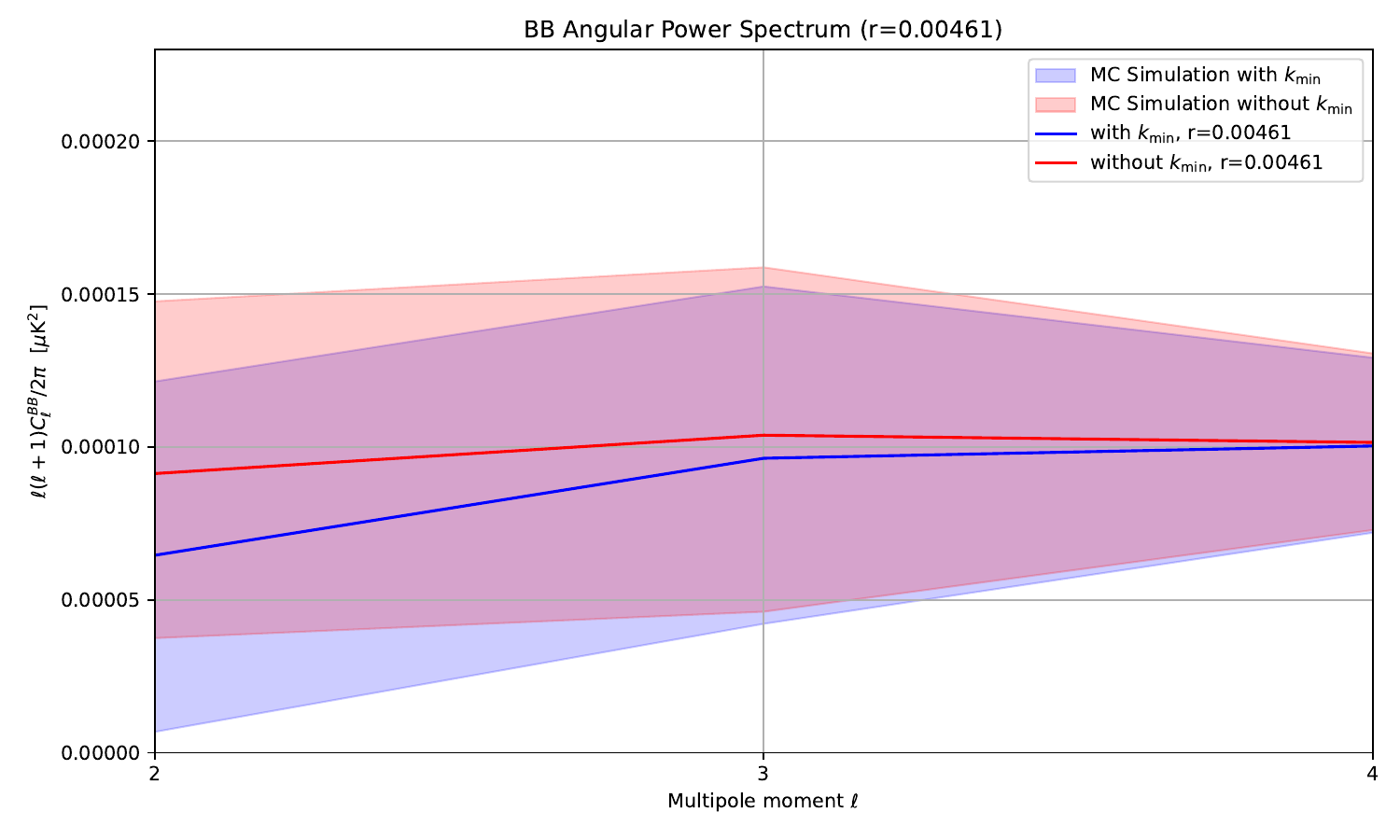}
\caption{BB angular power spectrum for $\ell \leq 4$. In all three panels, the blue 
and red lines show the results with and without $k_{\rm min}$. The shaded regions 
represent the expected total LiteBIRD errors, including foreground residuals. The 
top panel is for $r = 0.036$, including the forecast $\pm 1\sigma$ region; the middle
panel is for $r = 0.02$, with the expected $\pm 1\sigma$ region; the bottom panel
is for $r = 0.00461$, with the forecast $\pm 1\sigma$ region.}\label{fig:CL_BB}
\end{figure}

As we can see from Figure~\ref{fig:CL_BBvaryr}, similarly to the temperature and $E$-mode 
cases, the impact of $k_{\rm min}$ is primarily observed in the range $\ell \leq 4$. Therefore, 
we have magnified this portion of the BB angular power spectrum in Figure~\ref{fig:CL_BB} to 
highlight the impact of $k_{\rm min}$ and compare it to the accuracy of the LiteBIRD mission. 
Figure~\ref{fig:CL_BB} was produced through Monte Carlo (MC) simulations. We took the 
theoretically predicted values of $C_{\ell}$'s (which we show in Figure~\ref{fig:CL_BBvaryr}) 
and the predicted error bars (total errors) of the LiteBIRD mission \citep{LiteBIRD:2022}, 
expected to provide the most sensitive $B$-mode measurement, together with an assumption 
that these errors are Gaussian (as presented in LiteBIRD's publications). We then randomized 
the values of the predicted $C_{\ell}$'s based on these errors, and subsequently calculated 
the average and the uncertainty at each $\ell$.

From Figure~\ref{fig:CL_BB}, we can easily see how the tensor-to-scalar ratio $r$ will affect 
the detection of $k_{\rm min}$. For $r=0.00461$, it is almost impossible to distinguish between 
the two theoretical curves as they are completely submerged within each other's $1\sigma$ region. 
For $r=0.02$ and $r=0.036$, the two cases can be distinguished at $\ell=2$ and can be barely 
distinguished at $\ell=3$. At $\ell \geq 4$, they are not distinguishable. Since $0.036$ is 
already the upper limit of $r$, we are not optimistic about the detection of $k_{\rm min}$ 
in the $B$-mode angular power spectrum. However, this also means that even if upcoming missions 
such as LiteBIRD do not observe any low $\ell$ anomalies as seen in the temperature data, it 
does not necessarily mean that $k_{\rm min}$ does not exist, especially if the upper limit 
of $r$ is constrained to smaller values as more accurate measurements become available.

\begin{figure}[ht!]
\centering    
\includegraphics[width=\columnwidth]{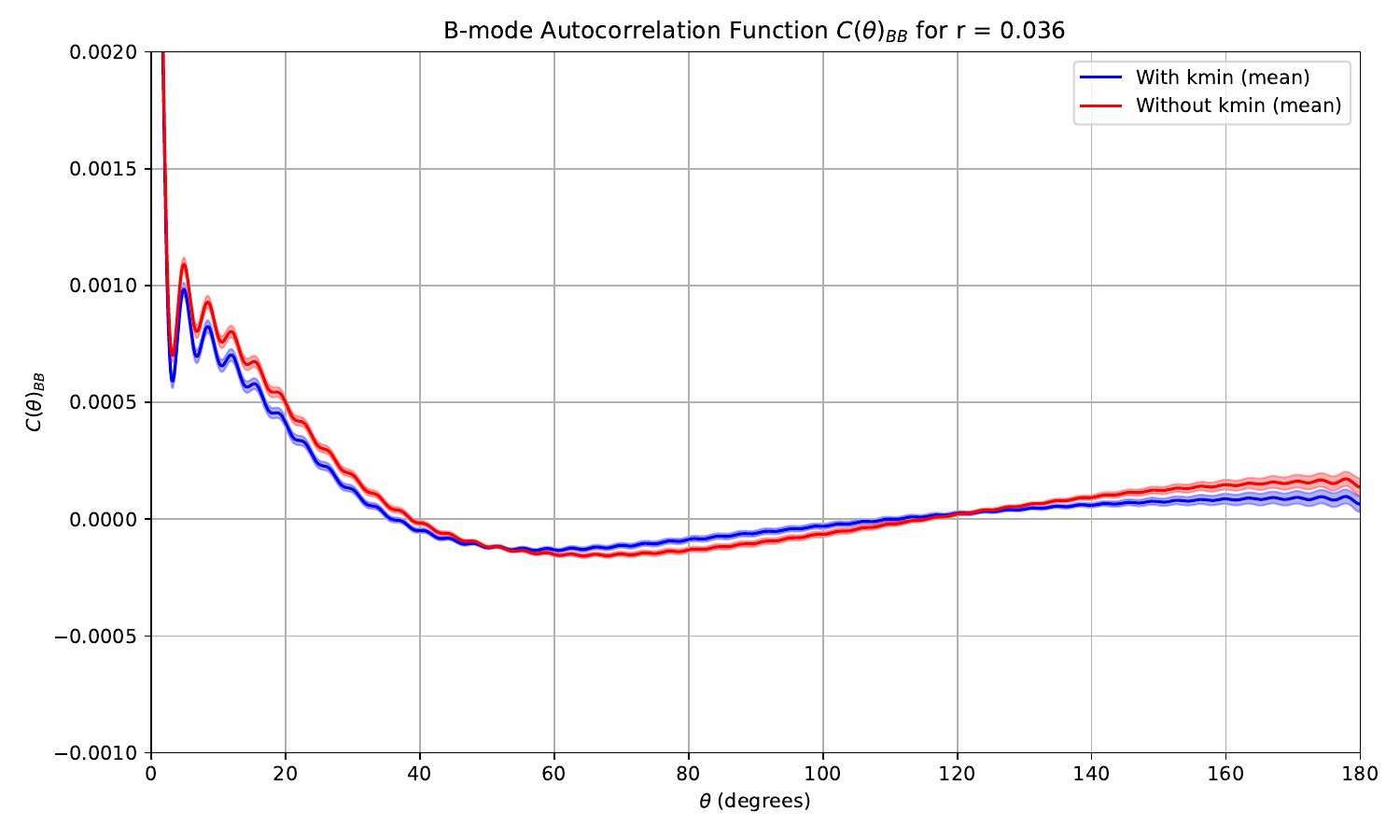}
\includegraphics[width=\columnwidth]{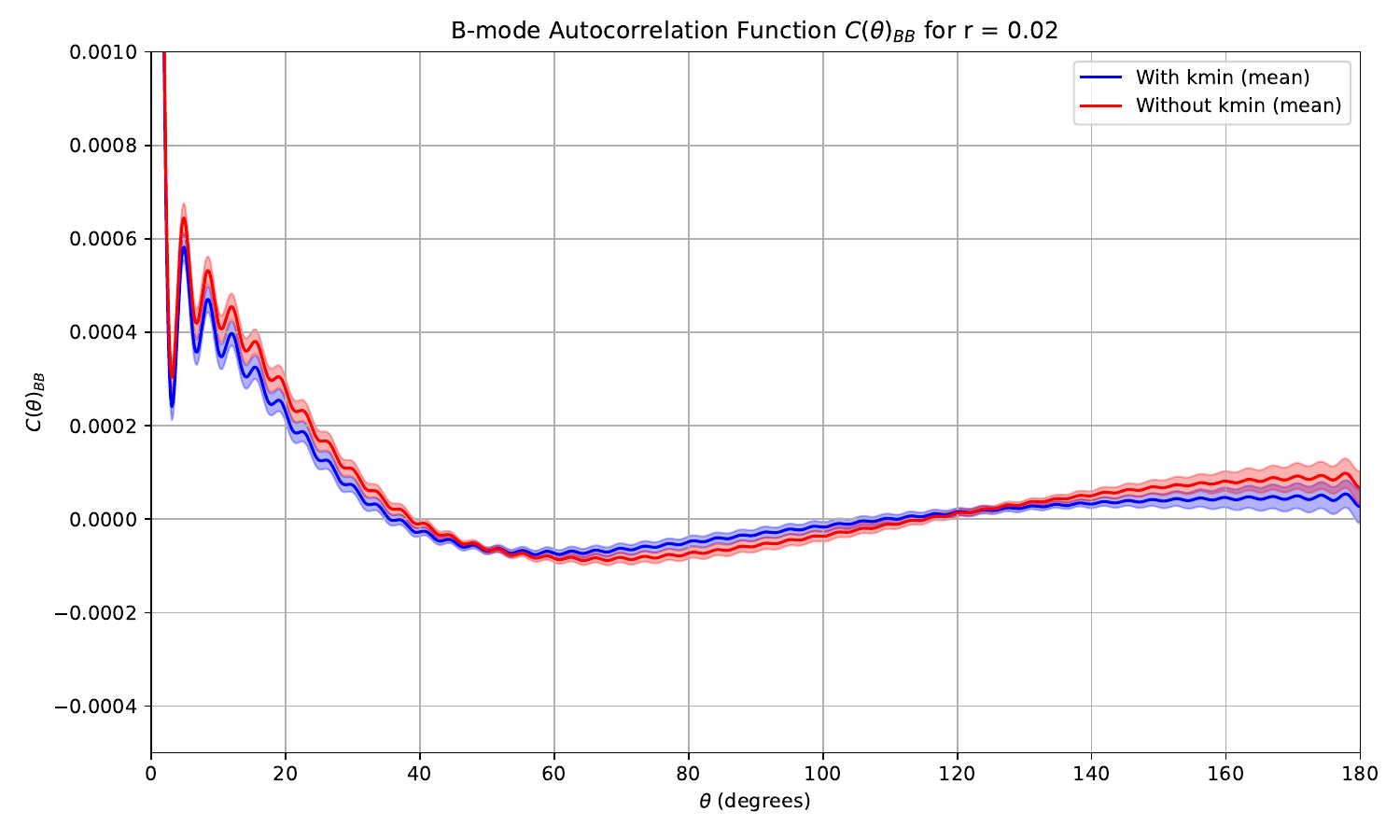}
\includegraphics[width=\columnwidth]{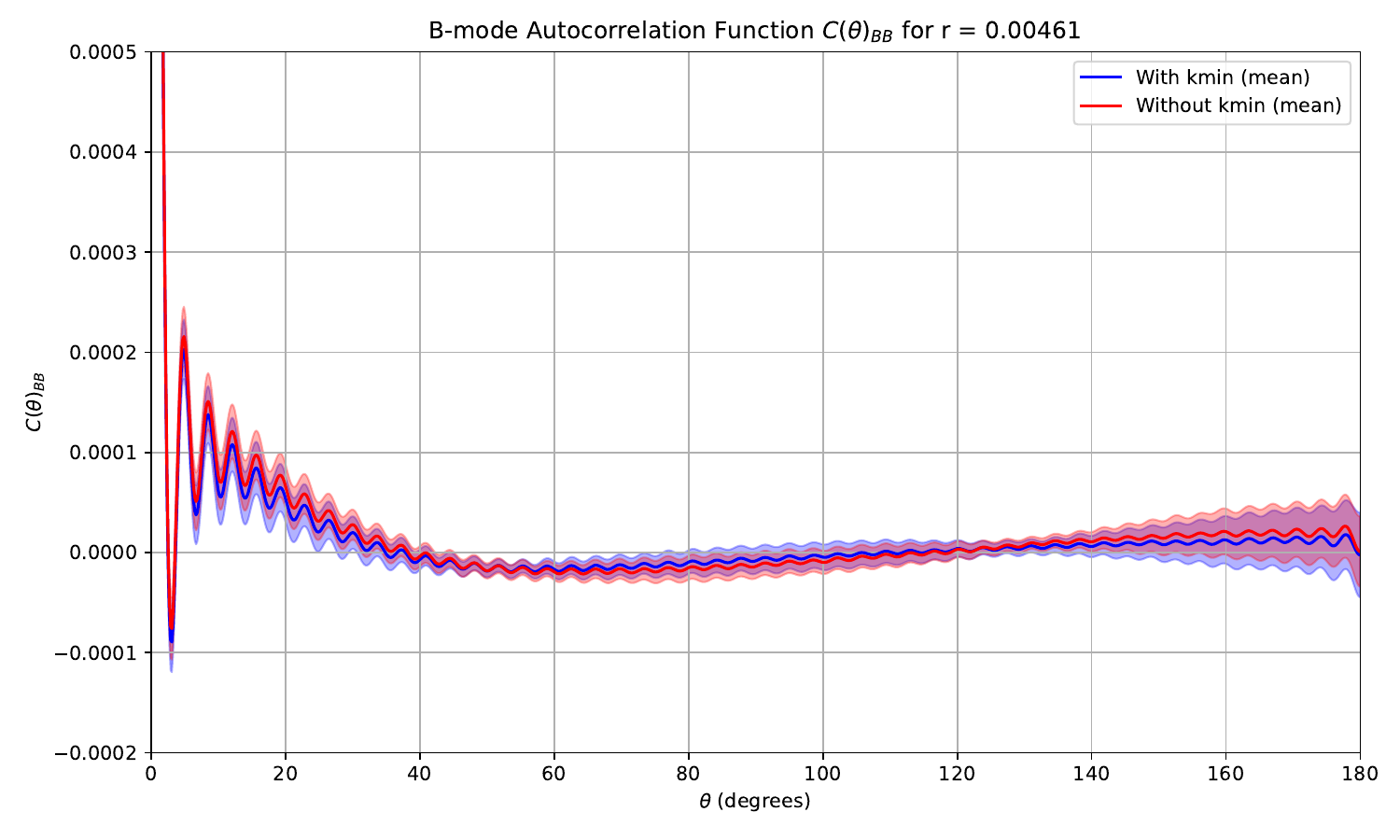}
\caption{$BB$ angular correlation function including $\ell < 100$ modes. In all three panels, 
the blue and red lines show the results with and without $k_{\rm min}$ and the corresponding 
$1\sigma$ region was generated via a MC simulation utilizing the predicted LiteBIRD's total 
errors. The top panel is for $r = 0.036$, including the forecast $\pm 1\sigma$ region; the 
middle panel is for $r = 0.02$, with the expected $\pm 1\sigma$ region; the bottom panel
is for $r = 0.00461$, with the forecast $\pm 1\sigma$ region.}\label{fig:Ctheta_BB}
\end{figure}

As was the case for the angular power spectrum, we shall calculate the $B$-mode angular 
correlation function for three different $r$ values: $r=0.036$, $r=0.02$, and $r=0.00461$. 
And similarly to the $E$-mode analysis, we use Equation~(\ref{CthetaEB}) to directly 
calculate the $B$-mode angular correlation function, whose results are shown in 
Figure~\ref{fig:Ctheta_BB}.

The three plots in this figure correspond to the three different values of $r$. In all 
three panels, the blue and red curves show the results with and without $k_{\rm min}$. 
The shaded regions represent the predicted $1 \sigma$ errors, which correspond to 
LiteBIRD's predicted total errors. The method we use to create these plots is similar 
to that applied to the $E$-mode analysis. We take the theoretically predicted $C_{\ell}^{BB}$'s 
of the different cases (with and without $k_{\rm min}$ for different values of $r$), and
then run a one thousand-step MC simulation on each of these sets. To be more specific, 
we randomize these six sets of $C_{\ell}^{BB}$'s according to LiteBIRD's predicted total 
errors. For each simulation, we calculate the angular correlation function, limiting our 
analysis to a certain range of $\ell$'s ($\ell < 100$) for two reasons. First, similarly 
to the $E$-mode case, including high $\ell$ terms makes the angular correlation function 
fuzzy. More importantly, on the angular power spectrum plots , the differences between 
the cases with and without $k_{\rm min}$ are only apparent at very low $\ell$'s, such 
as $\ell \leq 5$. Second, LiteBIRD provided total error predictions only for a limited 
range of $\ell$'s ($\ell < 200$). We then calculate the average and standard deviation 
for each of the six cases individually at all angles. On the plots, the theoretical 
curves represent the calculated averages, and the $1\sigma$ regions represent the 
standard deviations.

The results shown in Figure~\ref{fig:Ctheta_BB} are quite interesting. For the $r=0.00461$ 
case, the conclusion is similar to that of the angular power spectrum, where the difference 
between the cases with and without $k_{\rm min}$ is too small at all angles, and the two 
theoretical curves are well within each other's $1 \sigma$ region for most of the angles. 
For the other two values of $r$, however, the situation is quite different. In the angular 
power spectrum analysis, we concluded that even for $r=0.036$, where we observe the largest 
separation between the two cases, the difference is barely identifiable at $\ell=3$ and is 
only obvious at $\ell=2$. However, in the angular correlation function, for $r=0.036$, we 
can clearly see a distinct difference between the cases with and without $k_{\rm min}$ at 
all angles, especially for $\theta < 120^\circ$, where the two curves are far outside each 
other's $1 \sigma$ region. For the $r=0.02$ case, the two curves are outside each other's 
$1 \sigma$ region at most angles. Therefore, we believe that in terms of detecting 
$k_{\rm min}$ in the $B$-mode polarization, the angular correlation function is more 
helpful than the angular power spectrum.

We are currently still unable to confirm the existence of $k_{\rm min}$ with $B$-mode 
observations because there are no available data to use in the analysis. As we have shown 
above, however, if $B$-mode polarization is detected by the next-generation CMB missions, 
such as LiteBIRD, and the tensor-to-scalar ratio is not too far from its current upper limit, 
it should be possible to identify traces of $k_{\rm min}$ in the $B$-mode data, especially 
in the angular correlation function. If $B$-mode polarization continues to elude detection, 
and the tensor-to-scalar ratio continues to decrease as the accuracy of detectors improves, 
it is possible that the value of $r$ may fall below the threshold for us to detect 
$k_{\rm min}$. Of course, this does not mean $k_{\rm min}$ does not exist; it merely
means that when the tensor-to-scalar ratio is too small, as predicted by theory, the 
impact of $k_{\rm min}$ on $B$-mode is not as pronounced as its effect on the temperature 
and $E$-mode signals.

\subsection{Analysis of the Q+U and Q-U Modes}\label{QU}
Up to this point, we have examined the impact of $k_{\rm min}$ on the CMB's temperature and 
polarization signals by calculating the theoretical predictions of the angular power spectrum 
and angular correlation function for cases with and without this cutoff. And we then compared 
these results to the current {\it Planck} measurements, or described how the expected different
outcomes could be used with future missions, such as LiteBIRD, to confirm or reject the existence
of $k_{\rm min}$ in the primordial power spectrum. This approach of calculating the angular 
correlation function is straightforward, as one can see in Equations~(\ref{CthetaEB}) and 
(\ref{Ctheta_TE}). 

Given the potential importance of identifying a non-zero $k_{\rm min}$ to inflationary 
theory, we will next check these results by carrying out an independent analysis of the 
angular correlation function, based on an alternative approach. In this section, we adopt
the method used by CAMB and, following CAMB's convention, refer to these angular correlation 
functions as the $Q+U$ mode and $Q-U$ mode \citep{Lewis:2000}. Note, however, that they are 
not simply the sum or subtraction of the $Q$ and $U$ modes. 

As we discussed earlier, the $Q$ and $U$ modes are ill-defined when viewed on the entire
celestial sphere, because $Q$ and $U$ represent the linear polarization components along 
two orthogonal directions, and such a representation is not unique or invariant in a 
spherical coordinate system \citep{Zaldarriaga:1997}. Different points on the sphere have 
different local coordinate 
systems, causing the direction of linear polarization to change with position. This makes 
it difficult to define consistent $Q$ and $U$ modes across the entire sky. Additionally, 
$Q$ and $U$ components mix with each other under coordinate system rotation, further 
complicating their definition on the sphere.

However, by using the complex forms of the polarization components, $Q + iU$ and $Q - iU$, 
this problem can be overcome. As we already explained earlier, this complex form of 
representation is known as spin-weighted fields, where $Q + iU$ is a spin $+2$ field and 
$Q - iU$ is a spin $-2$ field. Spin-weighted fields are rotationally invariant in the 
spherical coordinate system, meaning that when the coordinate system rotates, the change 
in the spin-weighted field is represented only by a phase factor, rather than a mixing of 
components. Therefore, using $Q + iU$ and $Q - iU$ allows for consistent polarization 
modes to be defined across the entire celestial sphere, which is particularly effective 
in CMB polarization analysis. The $Q+U$ and $Q-U$ modes we introduced here are actually 
$Q+iU$ and $Q-iU$ modes.

To compute the angular correlation function, we use the spin-weighted spherical harmonics, 
which naturally incorporate the rotational properties of the celestial sphere. For example, 
the angular correlation function $C_{Q+U}(\theta)$ is given by \citep{NG:1999}
\begin{equation}
C_{Q+U}(\theta) = \sum_\ell \sqrt{\frac{2\ell + 1}{4\pi}} \left(C_\ell^{EE} + 
	C_\ell^{BB}\right) {_2Y_\ell^{-2}}(\theta, 0)\;.
\label{Ctheta_Q+U}
\end{equation}
Similarly, the angular correlation function $C_{Q-U}(\theta)$ is
\begin{equation}
C_{Q-U}(\theta) = \sum_\ell \sqrt{\frac{2\ell + 1}{4\pi}} \left(C_\ell^{EE} - C_\ell^{BB}\right)   
 {_2Y_\ell^2}(\theta, 0)\;.
\label{Ctheta_Q-U}
\end{equation}

In the CAMB software, these angular correlation functions are computed using the Wigner 
d-functions $d_{22}(\theta)$ and $d_{2,-2}(\theta)$. Specifically, CAMB uses the 
following formulas to compute the polarization correlation functions \citep{Lewis:2000}:
\begin{equation}
\begin{aligned}
C_{Q+U} (\theta) = \sum_{\ell=2}^{\ell_{\text{max}}} \left( \frac{2\ell+1}{4\pi} \right) 
	(C_\ell^{EE} + C_\ell^{BB}) d_{2,2} (\theta) \,,\\
C_{Q-U} (\theta) = \sum_{\ell=2}^{\ell_{\text{max}}} \left( \frac{2\ell+1}{4\pi} \right) 
	(C_\ell^{EE} - C_\ell^{BB}) d_{2,-2} (\theta)\,.
\end{aligned}
\label{Ctheta_Q+-U_CAMB}
\end{equation}
In CAMB's calculations, the Wigner $d$-functions are generated from the following equations:
\begin{equation}
\begin{aligned}
d_{2,2} = &\frac{\left[(4x - 8) + \ell(\ell + 1)\right] P_\ell(x) + 4 
	\left( \frac{1 - x}{1 + x} \right) (1 + x)}{(\ell + 2)(\ell - 1)} \notag \\
&+ \frac{4 \left( \frac{1 - x}{1 + x} \right) \left( \frac{x - 2}{\ell(\ell + 1)} \right) 
	P'_\ell(x)}{(\ell + 2)(\ell - 1)} \,,\\
d_{2,-2} = &\frac{\left[\ell(\ell + 1) - (4x + 8)\right] P_\ell(x) + 4 
	\left( \frac{1 + x}{1 - x} \right) (-(1 - x))}{(\ell + 2)(\ell - 1)} \notag \\
&+ \frac{4 \left( \frac{1 + x}{1 - x} \right) \left( \frac{x + 2}{\ell(\ell + 1)} \right) 
	P'_\ell(x)}{(\ell + 2)(\ell - 1)}\,,
\end{aligned}
\label{d22}
\end{equation}
where $x$ stands for $\cos(\theta)$.  These formulas leverage the properties of the Wigner 
d-functions to ensure accurate and efficient computation of the angular correlation function, 
taking into account the spin-$2$ nature of the polarization fields.

\begin{figure}
\centering
\includegraphics[width=\columnwidth]{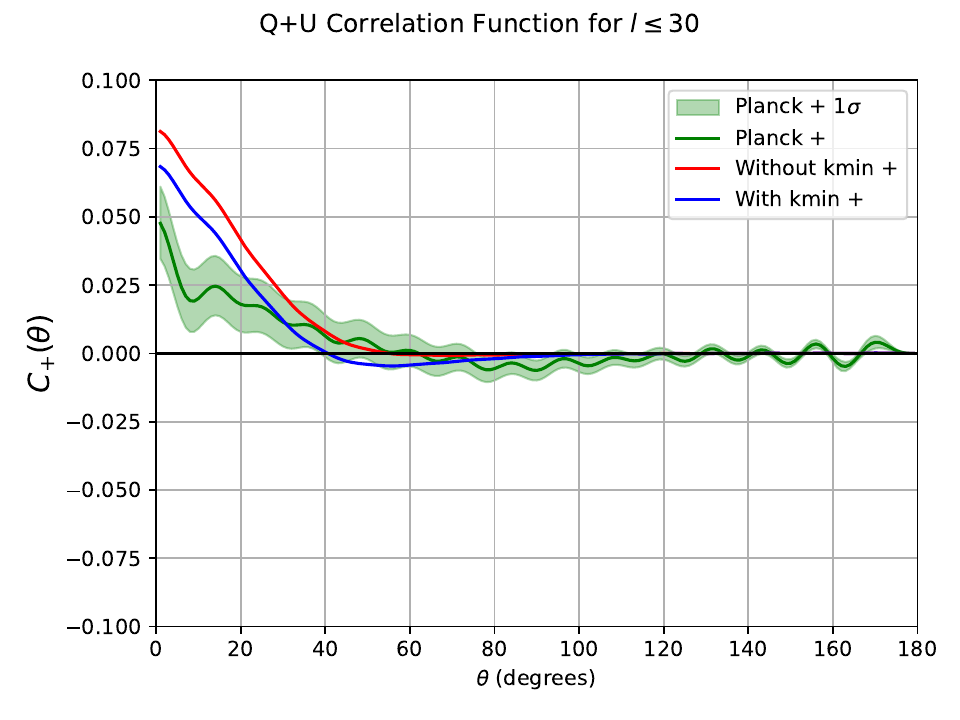}
\includegraphics[width=\columnwidth]{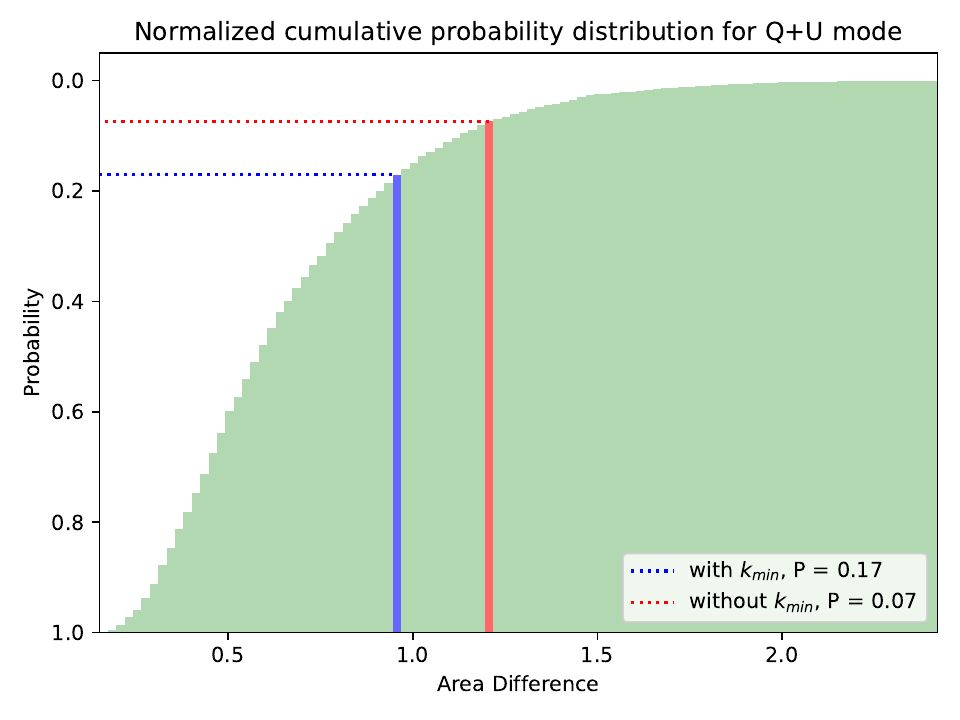}
\caption{$Q + U$ angular correlation function. Top: the $Q + U$ angular 
correlation function including $\ell \leq 30$ modes. The blue curve includes the effects 
of $k_{\text{min}}$, while the red curve represents calculations without $k_{\text{min}}$, 
i.e., the conventional application of the standard model. The green curve represents the 
angular correlation function calculated from the {\it Planck}-2018 data. The corresponding 
$1\sigma$ region was generated from a three thousand step MC simulation utilizing the total 
errors in {\it Planck}'s measurements. Bottom: The cumulative probability distribution of 
the $Q + U$ mode, generated from the area differences between the {\it Planck} curve and 
the three thousand realizations mentioned above, which were produced using {\it Planck}'s 
total errors. Blue represents the model with $k_{\rm min}$; red represents the model without 
$k_{\rm min}$.}\label{fig:Ctheta_Q+U}
\end{figure}

\begin{figure}[ht!]
\centering
\includegraphics[width=\columnwidth]{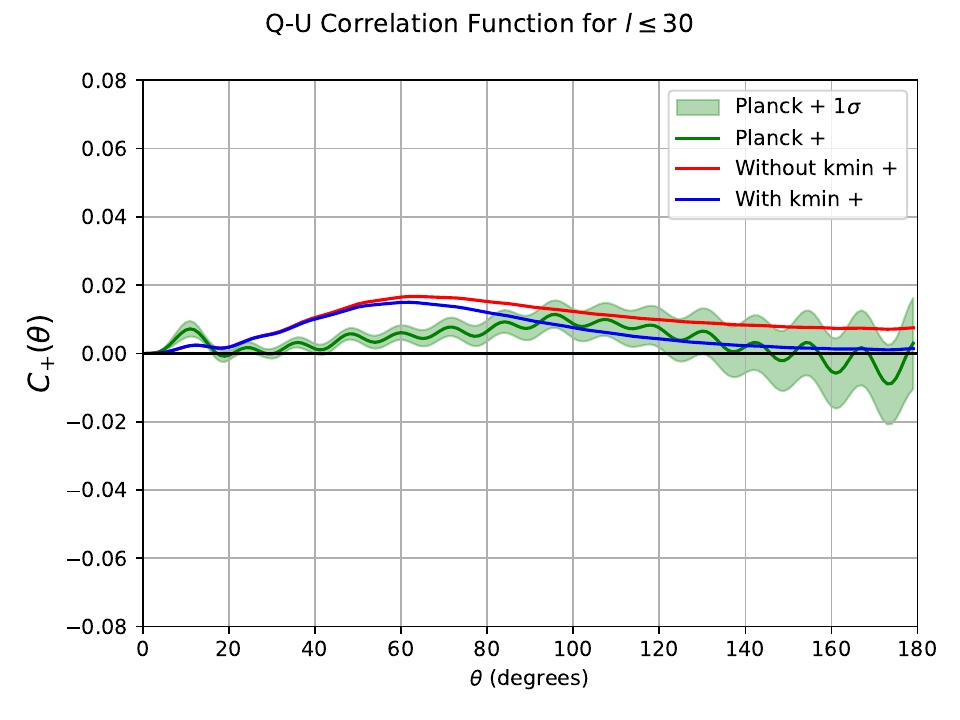}
\includegraphics[width=\columnwidth]{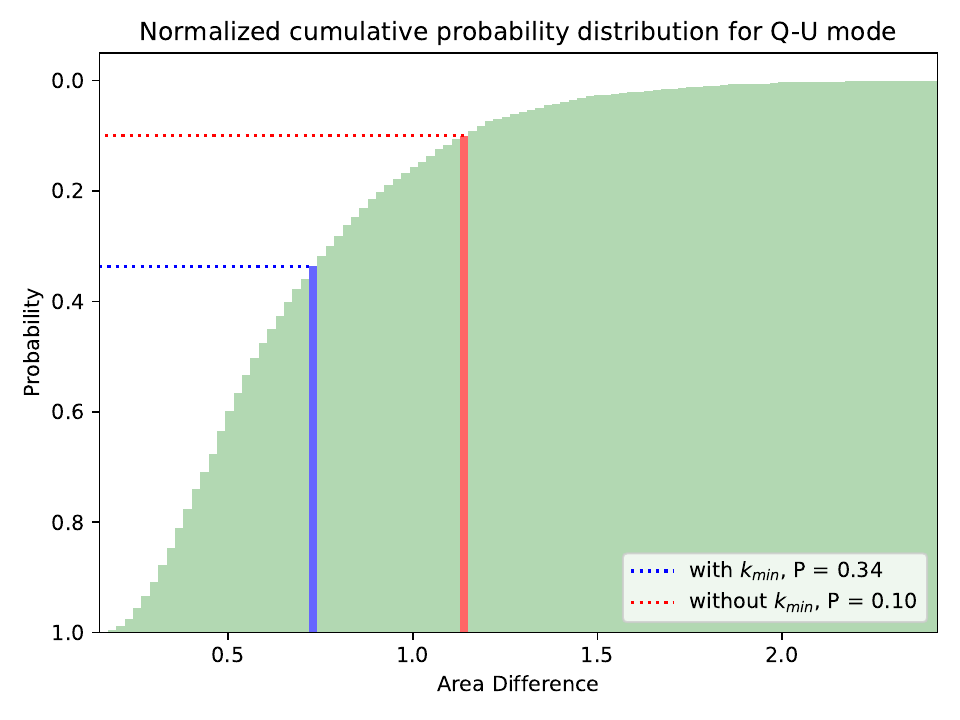}
\caption{$Q - U$ angular correlation functions. Top: the $Q - U$ angular 
correlation function including $\ell \leq 30$ modes. The blue curve includes the effects 
of $k_{\text{min}}$, while the red curve represents calculations without $k_{\text{min}}$, 
i.e., the conventional application of the standard model. The green curve represents the 
angular correlation function calculated from the {\it Planck}-2018 data. The corresponding 
$1\sigma$ region was generated from a three thousand step MC simulation utilizing the total 
errors in {\it Planck}'s measurements. Bottom: The cumulative probability distribution of 
the $Q - U$ mode, generated from the area differences between the {\it Planck} curve and 
the three thousand realizations mentioned above, which were produced using {\it Planck}'s 
total errors. Blue represents the model with $k_{\rm min}$; red represents the model without 
$k_{\rm min}$.}\label{fig:Ctheta_Q-U}
\end{figure}

The results of these calculations are shown in Figures~\ref{fig:Ctheta_Q+U} and 
\ref{fig:Ctheta_Q-U}. In both cases, the red curve represents the theoretical prediction 
without $k_{\rm min}$, while the blue curve shows outcome with a non-zero $k_{\rm min}$. 
The results calculated from {\it Planck}'s data are shown in green along with their 
$1 \sigma$ error region. For both sets of plots, we selected the range of 
multipoles to be $\ell < 30$, because $k_{\rm min}$ affects only the low-$\ell$ region 
of $C_\ell$. Additionally, including high-$\ell$ terms of $C_\ell$ makes the angular 
correlation function fuzzy. Thus, selecting this range allows us to better gauge the 
impact of $k_{\rm min}$.

All the theoretical curves in these two plots were calculated entirely with CAMB. 
Specifically, the theoretical prediction of the angular correlation function shown 
here is calculated from the angular power spectrum produced by CAMB. During this 
process, the most updated parameters \citep{PlanckVI:2020} from {\it Planck} were used. 
The {\it Planck} measurements shown in these figures were derived through 
three thousand steps of Monte Carlo (MC) simulations. Specifically, we took the 
{\it Planck} measurements of the $E$-mode angular power spectrum along with their total 
errors. We then randomized these $C_\ell$'s into three thousand sets of mock 
$C_\ell$'s. For each set of $C_\ell$'s, we calculated an angular correlation function 
using CAMB. We then calculated the mean and the standard deviation of the angular 
correlation function at each angle. One point to stress is that the evaluation of 
the $Q+U$ and $Q-U$ angular correlation functions requires both the $C_\ell^{EE}$'s 
and $C_\ell^{BB}$'s. {\it Planck} only has $E$-mode measurements, however. But since 
the tensor-to-scalar ratio has already been constrained to a very small value, 
$r < 0.036$, using only the $E$-mode angular power spectrum in these calculations is a 
very good approximation. Therefore, for both the theoretical predictions and the 
{\it Planck} measurements, we used only the $E$-mode angular power spectrum.

The $Q+U$ mode angular correlation function is plotted in Figure~\ref{fig:Ctheta_Q+U}. 
The top panel shows the results of the calculation using the $C_\ell$'s within the range 
$\ell \leq 30$. As we can see from the plot, both theoretical curves fit the measurements
very poorly and both miss the $1\sigma$ error region of the measurements at $\theta < 
20$ degrees. It is still clear, however, that introducing $k_{\rm min}$ brings the 
theoretical predictions closer to the measurements.

The $Q-U$ mode angular correlation function is plotted in Figure~\ref{fig:Ctheta_Q-U}. 
The top panel shows the results of the calculation using the $C_\ell$'s within the range 
$\ell \leq 30$. Although both theoretical curves fail to fit the measurements very 
well at $\theta < 60$ degrees, the prediction with $k_{\rm min}$ fits the measurements 
quite well at $\theta > 80$ degrees, whereas the curve without $k_{\rm min}$ fails 
significantly at all angles, it falls outside the $1\sigma$ 
error region of the measurement at almost all angles.

Analogously to the $E$-mode correlation and the $TE$ cross-correlation, we 
use a cumulative probability approach to estimate the likelihood that the 
area difference between a theoretical curve and the {\it Planck} curve arises solely from 
the total errors of the {\it Planck} measurements. The cumulative probability distribution 
plot of the $Q + U$ mode is presented in the bottom panel of Figure~\ref{fig:Ctheta_Q+U}. 
The cumulative probability distribution plot of the $Q - U$ mode is presented in the bottom 
panel of Figure~\ref{fig:Ctheta_Q-U}. 

The results are as follows: For the $Q + U$ mode, in the case without $k_{\rm min}$, the 
probability that the area difference is due to the total errors of the {\it Planck} 
measurements is 0.07. In the case with $k_{\rm min}$, the probability is 0.17. For the 
$Q - U$ mode, in the case without $k_{\rm min}$, the probability is 0.10. In the case 
with $k_{\rm min}$, the probability is 0.34.

These results for the $Q+U$ and $Q-U$ angular correlation functions clearly confirm the 
findings of previous studies and the results presented earlier in this paper. In both cases, 
introducing $k_{\rm min}$ helps bring the theoretical predictions closer to the measurements. 
This is especially true in the $Q-U$ case, in which the difference between the models
with and without $k_{\rm min}$ is significant, and the one with $k_{\rm min}$ is a much
better fit to the data. These results favor the model with $k_{\rm min}$ and thus add
to the justification for the existence of $k_{\rm min}$.

\section{Conclusion}\label{conclusion}
The hypothesized delayed initiation of inflation would produce a rigid cut-off, 
$k_{\rm min}$, for both the scalar and tensor primordial power spectra 
\citep{LiuMelia:2020,Liu:2024a}. This will impact all aspects of the CMB observations, 
including the temperature and polarization signals. {\it Planck} has conducted precise 
measurements of the temperature fluctuations and the $E$-mode polarization 
\citep{PlanckVI:2020}. $B$-mode polarization, however, has not been detected yet. 
One can anticipate that the next-generation missions, such as LiteBIRD, will be capable 
of detecting the $B$-mode and finally determine the tensor-to-scalar ratio, $r$
\citep{LiteBIRD:2022}. In this paper, we have calculated the theoretical predictions 
of the angular power spectrum and the angular correlation function for both cases, 
with and without $k_{\rm min}$. And we have compared the theoretical predictions 
with the available measurements. For the $B$-mode, we compared the differences in 
the expected fits with and without $k_{\rm min}$, and discussed the possibility that 
LiteBIRD may be able to distinguish between these two cases, depending on the value of $r$.

Based on our analysis, the existence of $k_{\rm min}$ is already on firm ground. Its
introduction significantly helps to align the theoretical predictions with the existing 
data. Our analysis not only confirms the results obtained in earlier studies
\citep{MeliaLopez:2018,Melia:2021b}, based primarily
on the temperature signal, but goes further in providing even stronger evidence for a
non-zero cutoff to the primordial power spectrum.

Specifically, the introduction of $k_{\rm min}$ enhances the theoretical fit of the $E$-mode 
polarization. The improvement is noticeable in both the angular power spectrum and the 
angular correlation function, for which the theoretical curves with $k_{\rm min}$ fit the
{\it Planck} data more closely compared to the standard model without the cutoff.

The $TE$ cross-correlation function analysis further supports the impact of $k_{\rm min}$. 
Its inclusion in the model reduces discrepancies between the theoretical predictions and 
the data across various angles, thereby easing tensions observed in the standard model.

Although current $B$-mode observations are very limited, our analysis indicates that 
$k_{\rm min}$ affects the $B$-mode angular power spectrum and angular correlation function
in measurable ways. The difference between models with and without $k_{\rm min}$ is more 
pronounced at low multipoles, however, so that ought to be an important focus of upcoming
observations, particularly if the tensor-to-scalar ratio $r$ is not too much smaller than
its current upper limit.

We have also confirmed these results with a separate analysis based on the use of
the $Q+U$ and $Q-U$ angular correlation functions, which fully support the conclusion
that introducing $k_{\rm min}$ mitigates the tension between the observations and the
current suite of theoretical predictions. This is especially evident in the $Q-U$ case, 
where the difference between the fits with and without $k_{\rm min}$ is significant. 

For all the cases ($TT$, $EE$, $TE$, $Q+U$, and $Q-U$) where we have 
calculated the angular correlation function of the predictions and the {\it Planck} 
measurements, we compared the two models (with and without $k_{\rm min}$) using a 
cumulative probability distribution plot. In all cases, the model with $k_{\rm min}$ 
shows a much higher likelihood that the discrepancy between theory and measurements 
is solely due to the total errors of the measurements. In other words, the model with 
$k_{\rm min}$ is favored in all cases.

In conclusion, our comprehensive analysis across temperature, $E$-mode, $TE$ 
cross-correlation, and $B$-mode polarization signals robustly supports the 
introduction of $k_{\rm min}$ in the primordial power spectrum. This cutoff 
not only helps resolve the large-angle anomalies observed in the CMB, but 
also provides a more accurate representation of the early Universe's inflationary 
epoch. Future observations, particularly of the $B$-mode polarization, will
be crucial for further validating the existence and implications of $k_{\rm min}$.

\begin{acknowledgments}
FM is grateful to Amherst College for its support through a John Woodruff Simpson Fellowship.
\end{acknowledgments}

\vfill\newpage
\bibliographystyle{aasjournal}
\bibliography{ms}{}

\end{document}